\documentclass[11pt,a4paper]{article}
\pdfoutput=1
\usepackage{jstyle}
\usepackage{amsmath, amsfonts, amssymb}
\usepackage{graphicx, hyperref, color, verbatim}
\usepackage[dvipsnames]{xcolor}
\usepackage{float}
\usepackage{caption}
\usepackage{subcaption}

\graphicspath{{fig/}}

\usepackage{tikz}
\usetikzlibrary{arrows,shapes}
\usetikzlibrary{trees}
\usetikzlibrary{matrix,arrows} 				% For commutative diagram
\usetikzlibrary{positioning}				% For "above of=" commands
\usetikzlibrary{calc,through}				% For coordinates
\usetikzlibrary{decorations.pathreplacing}  % For curly braces
\usepackage{pgffor}							% For repeating patterns

\usetikzlibrary{decorations.pathmorphing}	% For Feynman Diagrams
\usetikzlibrary{decorations.markings}
\tikzset{
	photon/.style={decorate, decoration={snake}, draw=red},
	electron/.style={draw=blue, postaction={decorate},
		decoration={markings,mark=at position .55 with {\arrow[draw=blue]{>}}}},
	gluon/.style={decorate, draw=black,
		decoration={coil,amplitude=4pt, segment length=4pt}} ,
	vector/.style={decorate, decoration={snake}, draw},
	provector/.style={decorate, decoration={snake,amplitude=2.5pt}, draw},
	antivector/.style={decorate, decoration={snake,amplitude=-2.5pt}, draw},
	fermion/.style={draw=black, postaction={decorate},
		decoration={markings,mark=at position .55 with {\arrow[draw=black]{>}}}},
	fermionbar/.style={draw=black, postaction={decorate},
		decoration={markings,mark=at position .55 with {\arrow[draw=black]{<}}}},
	fermionnoarrow/.style={draw=black},
	fermionnoarrowsoft/.style={draw=blue},
	scalar/.style={dashed,draw=black, postaction={decorate},
		decoration={markings,mark=at position .55 with {\arrow[draw=black]{>}}}},
	scalarbar/.style={dashed,draw=black, postaction={decorate},
		decoration={markings,mark=at position .55 with {\arrow[draw=black]{<}}}},
	scalarnoarrow/.style={dashed,draw=black},
	scalarnoarrowsoft/.style={dashed,draw=blue},
	electron/.style={draw=black, postaction={decorate},
		decoration={markings,mark=at position .55 with {\arrow[draw=black]{>}}}},
	bigvector/.style={decorate, decoration={snake,amplitude=4pt}, draw},
}

\tikzstyle{block} = [draw, rectangle, 
minimum height=3em, minimum width=6em]

\makeatletter
\DeclareFontFamily{OMX}{MnSymbolE}{}
\DeclareSymbolFont{MnLargeSymbols}{OMX}{MnSymbolE}{m}{n}
\SetSymbolFont{MnLargeSymbols}{bold}{OMX}{MnSymbolE}{b}{n}
\DeclareFontShape{OMX}{MnSymbolE}{m}{n}{
    <-6>  MnSymbolE5
   <6-7>  MnSymbolE6
   <7-8>  MnSymbolE7
   <8-9>  MnSymbolE8
   <9-10> MnSymbolE9
  <10-12> MnSymbolE10
  <12->   MnSymbolE12
}{}
\DeclareFontShape{OMX}{MnSymbolE}{b}{n}{
    <-6>  MnSymbolE-Bold5
   <6-7>  MnSymbolE-Bold6
   <7-8>  MnSymbolE-Bold7
   <8-9>  MnSymbolE-Bold8
   <9-10> MnSymbolE-Bold9
  <10-12> MnSymbolE-Bold10
  <12->   MnSymbolE-Bold12
}{}

\let\llangle\@undefined
\let\rrangle\@undefined
\DeclareMathDelimiter{\llangle}{\mathopen}%
                     {MnLargeSymbols}{'164}{MnLargeSymbols}{'164}
\DeclareMathDelimiter{\rrangle}{\mathclose}%
                     {MnLargeSymbols}{'171}{MnLargeSymbols}{'171}
\makeatother
\def\ma{\mathcal}
\def\mbb{\mathbb}
\def\ie{\begin{equation}\begin{aligned}}
\def\fe{\end{aligned}\end{equation}}
\newcommand{\te}{\text}

\title{Notes on flat-space limit of holographic defect correlators in position space}

\author[a]{Qi Chen,}
\author[b,c]{Yi-Xiao Tao,}
\author[d]{Xinan Zhou}

\affiliation[a]{Department of Physics, Tsinghua University, Beijing 100084, China}
\affiliation[b]{Department of Mathematical Sciences, Tsinghua University, Beijing 100084, China}
\affiliation[c]{Nordita, KTH Royal Institute of Technology and Stockholm University, Hannes Alf\'{v}ens v\"{a}g 12, SE-106 91 Stockholm, Sweden}
\affiliation[d]{Kavli Institute for Theoretical Sciences, University of Chinese Academy of Sciences, Beijing 100190, China}

\emailAdd{chenq20@mails.tsinghua.edu.cn}
\emailAdd{taoyx21@mails.tsinghua.edu.cn}
\emailAdd{xinan.zhou@ucas.ac.cn}

\abstract{We study the large AdS radius limit of correlation functions in holographic defect CFTs. For two-point functions of operators inserted away from the defect, we derive a position space formula relating a certain scaling limit of the correlators to the flat-space scattering form factors. We show that our position space prescription is equivalent to the flat-space limit formula recently conjectured in Mellin space
and also test our result in a few nontrivial theories.}

\begin{document}
\maketitle
\tableofcontents

\newpage

\section{Introduction}
Recently there has been significant progress in extending the bootstrap program for computing holographic correlators  \cite{Rastelli:2016nze,Rastelli:2017udc} to theories containing conformal defects \cite{Gimenez-Grau:2023fcy,Chen:2023yvw,Chen:2024orp,Zhou:2024ekb,Chen:2025yxg}. As was demonstrated in many examples over the last few years, the bootstrap strategy allows us to more effectively harness the computational power of AdS/CFT and was crucial in revealing the remarkable underlying simplicity (see \cite{Bissi:2022mrs} for a review). The recent extension to defects makes it possible to obtain new results at strong coupling in more interesting setups where superconformal symmetry is partially broken. The inclusion of defects also greatly enriches the study of holographic correlators by enlarging its scope in the dual bulk picture. The observables now include both scattering amplitudes in AdS and form factors of particles scattering with extended objects. It is an interesting open question how many of the hidden properties of AdS amplitudes \cite{Caron-Huot:2018kta,Rastelli:2019gtj,Giusto:2020neo,Behan:2021pzk,Alday:2021odx,Zhou:2021gnu,Alday:2022lkk,Aprile:2025kfk} can be extended to form factors. Moreover, the defect language also turns out to provide useful frameworks for studying certain systems which are usually not treated as defects due to their unconventional defect dimensions.\footnote{A conformal defect occupying the subspace $\mathbb{R}^p\subset \mathbb{R}^d$ breaks the conformal group $SO(d+1,1)$ to the subgroup $SO(p+1,1)\times SO(d-p)$. The conventional defects have dimensions $0\leq p\leq d-1$. However, it also makes sense to ``analytically continue'' $p$ to the unconventional dimensions $p=0,-1$, as long as the total dimension of the first $SO$ group factor is positive.} The efficiency of this defect perspective has recently been demonstrated in studying strongly coupled $\mathcal{N}=4$ super Yang-Mills theory on four dimensional real projective space \cite{Zhou:2024ekb} and correlators with heavy local operators such as giant gravitons \cite{Chen:2025yxg}. The defects in these examples have $-1$ and $0$ dimensions respectively. 

The study of holographic correlators in defect CFTs has so far mostly focused on the simplest supergravity limit. An important next step is to go beyond this limit and include the string theory and M-theory corrections. To compute these correlations, new information must be added as an input in addition to imposing the kinematic constraint required by superconformal symmetry. Similar to the defect-free case, two types of nontrivial constraints are particularly useful. The first type comes from non-perturbative techniques such as the supersymmetric localization which gives predictions for certain integrated correlators. These constraints have been studied for $\frac{1}{2}$-BPS line defects in 4d $\mathcal{N}=4$ SYM in \cite{Pufu:2023vwo,Billo:2023ncz,Dempsey:2024vkf,Billo:2024kri} and for giant graviton correlators in \cite{Brown:2024tru}. The second important condition arises from the consistency with the flat-space scattering form factor when the radius of the AdS space is taken to infinity. The focus of our paper is about this flat-space limit. The form factor in flat space provides powerful constraints by determining the leading behavior of the AdS counterpart in the high energy limit.\footnote{The input from flat space can also be imposed in a more refined way by assuming that the correlators arise from a wordsheet representation. This goes beyond the leading order and holds also for the curvature corrections. For four-point functions in defect-free CFTs, this idea has been extremely powerful. See, e.g., \cite{Alday:2023jdk,Alday:2023mvu,Alday:2023pzu,Alday:2024xpq,Alday:2024rjs,Wang:2025pjo,Alday:2025bjp} for various aspects.} However, it should be noted that in contrast to the defect-free case \cite{Okuda:2010ym,Penedones:2010ue}, how to take the flat-space limit of holographic defect correlators remains unclear and has not been analyzed in the literature. A first step in this direction was recently taken in \cite{Alday:2024srr} where a prescription in Mellin space was conjectured. The conjecture was based on the analysis of Witten diagrams and was checked in a few nontrivial examples. The goal of this work is to close this gap and give a rigorous derivation for this prescription. 

We will achieve our goal by performing a careful analysis of scattering in AdS and studying the flat-space limit of defect correlators in position space. For concreteness, we will focus on the case of two-point functions of local operators inserted away from the defect. We find that the flat-space limit of the holographic defect correlator is related to a particular scaling limit in the Lorentzian signature. In this limit, we need to zoom in around the kinematic configuration where a certain cross ratio is vanishing
\begin{equation}
    \Theta=-\frac{((x_1^a-x_2^a)^2+(|x_1^i|+|x_2^i|)^2)((x_1^a-x_2^a)^2+(|x_1^i|-|x_2^i|)^2)}{4|x_1^i|^2|x_2^i|^2}\to 0\;,
\end{equation}
but the ratio $\eta^2=-2\Theta R^2/\ell_s^2$ is kept fixed. Here $R$ is the radius of AdS and $\ell_s$ is an intrinsic length scale of the bulk physics. The indices $a$ and $i$ label respectively the directions parallel and orthogonal to the defect. This configuration can be realized by positioning the two operators in such a way that there is a point on the defect which is null separated from both the inserted operators as well as their reflection images with respect to the defect. The relevant scaling limit of the correlator is 
\begin{equation}
    \mathcal{G}_{\rm f.s.}(\eta,\chi)=\lim_{R/\ell_s\to\infty}\left(\frac{R}{\ell_s}\right)^{-(\Delta_1+\Delta_2+p+1-d)}\mathcal{G}\left(\Theta=-\frac{\eta^2\ell_s^2}{2R^2},\chi\right)\;,
\end{equation}
where 
\begin{equation}
\mathcal{G}(\Theta,\chi)=|x_1^i|^{\Delta_1}|x_2^i|^{\Delta_2}\llangle \mathcal{O}_{\Delta_1}(x_1)\mathcal{O}_{\Delta_2}(x_2)\rrangle\;,    
\end{equation}
is the defect correlator expressed as a function of cross ratios with 
\begin{equation}
    \chi=\frac{2x_1^jx_2^j}{|x_1^i||x_2^i|}\;,
\end{equation}
being the other independent conformal cross ratio. We will assume that  the flat-space limit of the correlator receives contributions only from a region localized near an AdS$_{p+1}$ subspace which is the bulk dual of the defect. We can then show by saddle point analysis that the dimensionless flat-space form factor $\mathcal{F}_{\rm flat}$ is related to the scaling limit of the correlator via an integral transform
\begin{equation}
\begin{split}
    \mathcal{G}_{\rm f.s.}(\eta,\chi)=&\frac{(-i)^{\Delta_1+\Delta_2-1}\pi^{\frac{p+1}{2}}}{\Gamma(\Delta_1)\Gamma(\Delta_2)2^{\frac{\Delta_1+\Delta_2-p-2}{2}}}\int_0^{\infty}du u^{\Delta_1+\Delta_2-2}(u\eta)^{\frac{1-p}{2}}\\
    &\times K_{\frac{p-1}{2}}(u\eta)\ma{F}_{\te{flat}}\left(S=-\frac{2+\chi}{\ell_s^2}u^2,Q=\frac{2u^2}{\ell_s^2}\right)\;.
\end{split}
\end{equation}
Here $Q=-(\vec{k}_1^{\parallel})^2=-(\vec{k}_2^{\parallel})^2$, $S=-\vec{k}_1\cdot \vec{k}_2$ are the Mandelstam variables in flat space in the 1-to-1 scattering process with a defect. Translating this formula into Mellin space, we can then prove the conjecture in \cite{Alday:2024srr}. Note that this position space formula is particularly useful when the Mellin representation is not well defined. We also emphasize that this result is valid for a wide range of defect dimensions $p=0,1,\ldots,d-1$. 

The rest of this paper is organized as follows. In Section \ref{Sec:reviewdefectfree}, we review the derivation in \cite{Okuda:2010ym} of the flat-space limit formula in defect-free CFTs as a warm up. In Section \ref{Sec:defectflatspacelimit}, we derive the defect version of the flat-space limit formula. We discuss in detail in Section \ref{Sec:analconti} the analytic continuation needed to apply the flat-space limit formula to Euclidean correlators. The equivalence with the Mellin space formula is proven in Section \ref{Sec:equivwithMellin}. In Section \ref{Sec:explicitexamples}, we apply the position space formula to a few nontrivial examples of tree-level holographic defect correlators and compare with the known results in flat space. These includes Wilson loops in $\mathcal{N}=4$ SYM, surface defects in 6d $(2,0)$ theories and giant graviton correlators. Finally, we conclude in Section \ref{Sec:discussions} with a brief discussion of future directions.

\section{Review of the defect-free case}\label{Sec:reviewdefectfree}
In this section, we review the derivation in \cite{Okuda:2010ym} which led to a flat-space limit formula for four-point functions in defect-free CFTs. As we will see, the defect CFT setup shares a lot of similarities. Therefore, it is useful to review the key ideas and techniques in detail. 

We consider an AdS$_{d+1}$ space with radius $R$. This can be conveniently described in terms of the embedding space coordinates $X\in \mathbb{R}^{2,d}$
\begin{equation}
X^2\equiv -(X^{-1})^{2}-(X^0)^2+(X^1)^2+\cdots+(X^d)^2=-R^2.
\end{equation}
Local operators are inserted at the boundary of the AdS space where each point is identified with a null ray in the embedding space
\begin{equation}
    P^2=0\;,\quad P\sim \lambda P\;.
\end{equation}
We focus on the following four-point function
\begin{equation}
	A(P_i)=\int_{\te{AdS}_{d+1}}\prod_{i=1}^{4}dX_i G_{B\partial}^{\Delta_i}(X_i,P_i)G(X_1,\cdots,X_4)\;.
\end{equation}
Here $G_{B\partial}^\Delta$ are the bulk-to-boundary propagators
\begin{equation}
	G_{B\partial}^\Delta(P,X)=\frac{1}{R^{(d-1)/2}}\frac{1}{(-2P\cdot X/R+i\epsilon)^{\Delta}}\;.
\end{equation}
The factor $ G(X_1, \cdots, X_4) $ is a correlation function in AdS$_{d+1}$. The holographic correlator is therefore constructed in a way similar to the Lehmann-Symanzik-Zimmermann (LSZ) reduction in flat space.

To properly analyze the flat-space limit, we will assume that the bulk theory has an intrinsic length scale $ \ell_s $ that is independent of the AdS radius $ R $. Although we have used the symbol $\ell_s$, the bulk theory does not need to be a string theory and $\ell_s$ can be other quantities than the string length. We are interested in the limit $\ell_s\ll R$ where the flat-space physics dominates. Therefore, we can ignore curvature effects of AdS and dominant contriubitions come from the region with $|X_i-X_j|\ll R$. The four-point function is then approximated as
\begin{equation}\label{eq:Aflatappro}
	A\approx\int_{\te{AdS}_{d+1}}dX_1 G_{B\partial}^{\Delta_1}(X_1,P_1)\int_{\mbb{M}}\prod_{i=2}^4 dY_iG_{B\partial}^{\Delta_i}(X_1+Y_i,P_i)G_{\te{flat}}(0,Y_2,Y_3,Y_4)\;,
\end{equation}
where $\mathbb{M}$ is a $d+1$ dimensional Minkowski space parameterized by $Y_i=X_i-X_1$. The AdS correlator $ G(X_1, \cdots, X_4) $ reduces to the flat-space amputated correlator in this limit. To proceed, we rewrite the bulk-to-boundary propagators using the Schwinger parameterization
\begin{equation}
	G^\Delta_{B\partial}(X,P)=\frac{(-i)^{\Delta}R^{\Delta}}{\Gamma(\Delta)R^{(d-1)/2}\ell_s^{\Delta}}\int_0^{\infty}\frac{d\beta}{\beta}\beta^{\Delta}e^{-2i\beta P\cdot X/\ell_s}\;.
\end{equation}
This makes it straightforward to perform the $Y$ integrals which gives the flat-space off-shell scattering amplitude in momentum space
\begin{equation}
	\int_{\mbb{M}}\prod_{i=2}^4d Y_ie^{-2i\beta_i P_i\cdot Y_i/\ell_s}G_{\te{flat}}(0,Y_2,Y_3,Y_4)=T^{(d+1)}\Big(k_1=-\sum_{i=2}^4 k_i,k_i=-2\beta_iP_i/\ell_s\Big)\;.
\end{equation}
We can then perform the integral over $X_1$. Let us define the combination $Q=\sum_{i=1}^4 \beta_i P_i$. Then the $X_1$ integral takes the form
\begin{equation}
    I_{X_1}=\int_{{\rm AdS}_{d+1}} dX_1 e^{-2i Q\cdot X_1/\ell_s}\;.
\end{equation}
The integral can be straightforwardly performed in the Poincar\'e coordinates
\begin{equation}
	\begin{split}
		I_{X_1}&=R^{d+1}\int \frac{dw_0 d^d\vec{w}}{w_0^{d+1}}e^{i\frac{\sum_i\beta_i R}{\ell_s}\frac{(w_0^2+(\vec{w}-\vec{x}_i)^2)}{w_0}}\\
		&=R^{d+1}\int\frac{dw_0}{w_0^{d+1}}e^{i\frac{\sum_i\beta_i R w_0}{\ell_s}}e^{-i\frac{R}{\ell_s}\frac{Q^2 }{\sum_i\beta_i w_0 }}\Big(\frac{\pi w_0 \ell_s}{-i\sum_i\beta_i R}\Big)^{\frac{d}{2}}\\
		&=\pi^{\frac{d}{2}}R^{d+1}\int_0^{\infty}\frac{dy}{y}(-iy)^{-\frac{d}{2}}e^{iy}e^{-i\frac{Q^2R^2}{\ell_s^2y}}\;,
	\end{split}
\end{equation}
where the change of variable $y=\frac{w_0R\sum_i\beta_i}{\ell_s}$ was used in the last step. This allows us to rewrite \eqref{eq:Aflatappro} as
\begin{equation}\label{eq:Aintegralybeta}
	\begin{split}
		A&\approx\pi^{\frac{d}{2}}R^{3-d}\prod_{i=1}^4\frac{(-i)^{\Delta_i}R^{\Delta_i}}{\Gamma(\Delta_i)\ell_s^{\Delta_i}}\int_0^{\infty}\frac{dy}{y}(-iy)^{-\frac{d}{2}}e^{iy}\\
		&~~~\times\int_0^{\infty}\prod_{i=1}^4\frac{d\beta_i}{\beta_i}\beta_i^{\Delta_i}e^{-i(\frac{R}{\ell_s})^2Q^2/y}T^{(d+1)}\Big(k_1=-\sum_{i=2}^4 k_i,k_i=-2\beta_iP_i/\ell_s\Big)\;.
	\end{split}
\end{equation}
An important point to notice is the large $(R/\ell_s)^2$ factor in the exponent. This allows us to use the saddle point approximation to perform the $\beta_i$ integrals. To do this, we need to diagonalize the real symmetric matrix $P_{ij}=-2P_i\cdot P_j$ in $Q^2=-\frac{1}{2}\sum_{i,j}\beta_i\beta_j P_{ij}$. Generically, the matrix has order 1 eigenvalues which render the integral exponentially suppressed. However, when one of the eigenvalues is very small, i.e., of order $(\ell_s/R)^2$, then the integral is enhanced. This is precisely the flat-space limit which we are interested in.  This condition can be achieved by taking the $P_i$ to be  
\begin{equation}\label{configPi}
	\begin{split}
		P_1&=(0,-1,-1,0,\vec{0})+\cdots,\\
		P_2&=(0,-1,1,0,\vec{0})+\cdots,\\
		P_3&=(0,1,\cos\theta,\sin\theta,\vec{0})+\cdots,\\
		P_4&=(0,1,-\cos\theta,-\sin\theta,\vec{0})+\cdots,
	\end{split}
\end{equation}
where we have used the global coordinates $P=(\cos\tau,\sin\tau,\bf{e})$ and $\bf{e}$ is a $d$ dimensional unit vector. The unit vector is further split into two parts where $\vec{0}$ represents the origin of $\mathbb{R}^{d-2}$. The small deviations away from this configuration are denoted by $\ldots$. We can then diagonalize the matrix $P_{ij}$ with the following linear combinations 
\begin{equation}\label{betai}
	\begin{split}
		\beta_1&=\frac{1}{2}(v-v_1-v_2+v_3)\;,~~~\beta_2=\frac{1}{2}(v-v_1+v_2-v_3)\;,\\
		\beta_3&=\frac{1}{2}(v+v_1-v_2-v_3)\;,~~~\beta_4=\frac{1}{2}(v+v_1+v_2+v_3)\;.
	\end{split}
\end{equation}
The corresponding eigenvalues are 
\begin{equation}
    \begin{split}
        {}&\lambda_0=0+\ldots\;,\quad \lambda_1=8+\ldots\;,\\
        {}&\lambda_2=-8\psi+\ldots\;,\quad \lambda_3=-8(1-\psi)+\ldots\;,
    \end{split}
\end{equation}
where we have defined
\begin{equation}
    \psi=\sin^2\left(\frac{\theta}{2}\right)\;.
\end{equation}
This recasts the $\beta$-integrals into the following decoupled form
\begin{equation}\label{intbeta}
	\begin{split}
	I_{\beta}&=\int\frac{dv}{\beta_1}\frac{dv_1}{\beta_2}\frac{dv_2}{\beta_3}\frac{dv_3}{\beta_4}\beta_1^{\Delta_1}\beta_2^{\Delta_2}\beta_3^{\Delta_3}\beta_4^{\Delta_4}e^{-i(-\frac{\lambda_0R^2}{2\ell_s^2})\frac{v^2}{y}}e^{-i(-4\frac{R^2}{\ell_s^2})\frac{v_1^2}{y}}\\
	&~~~\times e^{-i(4\psi\frac{R^2}{\ell_s^2})\frac{v_2^2}{y}}e^{-i[4(1-\psi)\frac{R^2}{\ell_s^2}]\frac{v_3^2}{y}}T^{(d+1)}(-t/s=\psi,s \ell_s^2=-4v^2)\;.
	\end{split}
\end{equation}
Here it is important to notice that the small deviations regularize the zero eigenvalue into a small $\lambda_0$. The Mandelstam variables are defined as $s=(k_1+k_2)^2$, $t=(k_1+k_4)^2$. A subtle point concerning the amplitude $T^{(d+1)}$ is that the external momentum $k_1$ in (\ref{eq:Aintegralybeta})  was off-shell. But here we have used the Mandelstam variables which are only valid when all particles are on-shell. The reason why this is allowed is that when we take the flat-space limit $k_1$ automatically becomes on-shell as well. More precisely, the square of $k_1$ is
\begin{equation}
    k_1^2\propto \sum_{i,j=2}^4 \beta_i\beta_j P_{ij}\;\;,
\end{equation}
which depends on the operator insertions as well as the Schwinger parameters. We can use the saddle point approximation for the non-flat directions $v_1$, $v_2$, $v_3$ in (\ref{intbeta}) and it is clear that the dominant contributions come from $v_1=v_2=v_3=0$. It follows from (\ref{betai}) that all $\beta_i$ are approximately equal to $\frac{1}{2}v$. Note that $v$ is not localized to zero in this limit and this ensures that the momenta are not  vanishing. It is then straightforward to verify using (\ref{configPi}) that $k_1$ is on-shell. Performing the saddle point integrals over $v_1$, $v_2$ and $v_3$, we get
\begin{equation}
	\begin{split}
		I_{\beta}=\frac{i\pi^{\frac{3}{2}}}{2^{2+\sum_i\Delta_i}}\Big(\frac{\ell_s}{R}\Big)^3\int_0^{\infty}\frac{dv}{v}v^{-3+\sum_i\Delta_i}\frac{(-i y)^{\frac{3}{2}}}{\sqrt{\psi(1-\psi)}}e^{-i\zeta^2\frac{v^2}{y}}\;,~~~\zeta^2=-\frac{\lambda_0R^2}{2\ell_s^2}\;.
	\end{split}
\end{equation}
Here $\lambda_0$ combines with $\ell_s/R$ to form a new scaling variable $\zeta$ which is kept finite when taking the flat-space limit. The integral over $y$ can then be performed which gives rise to the modified Bessel function
\begin{equation}\label{fsformula4pt}
\begin{split}
    A\approx&\frac{4(2\pi)^{\frac{3+d}{2}}\ell_s^3}{R^d}\prod_{i=1}^4\frac{(-i)^{\Delta_i}R^{\Delta_i}}{\Gamma(\Delta_i)(4\ell_s)^{\Delta_i}}\int_0^{\infty}\frac{dv}{v}v^{\sum_i\Delta_i-3}(\zeta v)^{\frac{3-d}{2}}\\
    &\times K_{\frac{3-d}{2}}(\zeta v)iT^{(d+1)}(-t/s=\psi,s \ell_s^2=-v^2)\;,
\end{split}
\end{equation}
where we have rescaled $v\to\frac{v}{2}$ to match the convention used in \cite{Okuda:2010ym}.

\section{Flat-space limit of defect correlators in position space}
\subsection{Flat-space limit formula}\label{Sec:defectflatspacelimit}
We now consider the defect case and focus on two-point functions of local operators inserted away from the defect. Our goal in this subsection is to derive a flat-space limit formula in position space which is analogous to (\ref{fsformula4pt}). To be precise, we will consider a ``probe brane'' setup where the $p$ dimensional defect in the CFT is dual to an AdS$_{p+1}$ subspace in AdS$_{d+1}$ with localized degrees of freedom coupled to the bulk. Although such defects are in a sense ``perturbative'' (when the defect is not heavy enough to back-react to the bulk geometry), they already cover many interesting physical setups as we will see in the examples in Section \ref{Sec:explicitexamples}. We will follow a similar logic as in the previous section to derive the flat-space formula, and the central result of this subsection is  (\ref{defectfslimit}). It will also be clear from the derivation that the flat-space limit formula is valid for a wide range of defect dimensions $0\leq p\leq d-1$.\footnote{One can also make sense of the case with $p=-1$ from the defect perspective. For example, it can be realized by placing a theory on real projective space \cite{Zhou:2024ekb}. However, such $(-1)$-dimensional defects are a bit special and the flat-space limit of defect correlators is not covered by our formula.}

Let us start with a generic defect two-point function in such a setup which can be written in the following form
\begin{equation}\label{defF2pt}
	G(P_1,P_2)=\int_{\te{AdS}_{d+1}}dX_1\int_{\te{AdS}_{d+1}}dX_2G_{B\partial}(P_1,X_1)G_{B\partial}(P_2,X_2)G(X_1,X_2)\;.
\end{equation}
Here $G(X_1,X_2)$ is a bulk two-point function in the presence of a holographic defect. Similar to the defect-free four-point correlator, the defect two-point function is also defined in the sense of an AdS LSZ reduction. Let us also comment that the defect two-point function has the physical meaning of a form factor in AdS describing the scattering of particles scattering with an extended object. We will assume that the AdS theory is governed by an intrinsic scale $\ell_s$ and the flat-space limit receives dominant contributions from the near-flat regions with $|X_i-X_j|\ll R$. Note that when there was no defect, we integrated over a flat Minkowski space around a bulk point which is then followed by integrating this bulk point over the whole AdS$_{d+1}$ space. The reason that we needed to integrate over the full space is that all points in AdS are equivalent and are related by isometry. However, an important difference here is that the presence of the defect breaks some of the bulk isometries. Furthermore, we will make the assumption that interesting physics happens near the AdS$_{p+1}$ brane within distances of order $\ell_s$. Therefore, the integration over equivalent reference points around which we need to zoom in should now only be with respect to AdS$_{p+1}$. More precisely, let us choose $X_1$ to be the reference point and write $X_2$ as $X_1+Y$ with $Y$ parameterizing the flat space $\mathbb{M}$. In the flat-space limit, the two-point function can be approximated by
\begin{equation}
	G\approx \int_{\te{AdS}_{p+1}}dX_1^{\parallel}\int_{\mbb{M}}dX_1^{\perp}dY^{\parallel}dY^{\perp}G_{B\partial}^{\Delta_1}(P_1,X_1)G_{B\partial}^{\Delta_2}(P_2,X_1+Y)G_{\text{flat}}(X_1^\perp,Y^{\parallel},Y^{\perp}).
\end{equation}
Here the AdS$_{p+1}$ defect splits $Y$ into a parallel part $Y^\parallel$ and a transverse part $Y^\perp$ (the meaning will be specified more precisely in a moment) and we need to integrate over both. We also need to integrate over the transverse part $X_1^\perp$ near the defect which is approximately flat under the assumption that the flat-space limit zooms in to the AdS defect. Note that the flat-space correlator depends on both $X_1^\perp$ and $Y^\perp$ but only on $Y^\parallel$. This is because translation symmetry is broken in the transverse directions but is preserved along the defect. Finally, we need to integrate over the parallel components $X_1^\parallel$ which form an AdS$_{p+1}$ subspace. 

To proceed, we follow a similar procedure as in the defect-free case. It is also convenient to rewrite the propagators as exponentials by using the Schwinger parameterization
\begin{equation}
		G_{B\partial}^\Delta(X,P)=\frac{(-i)^{\Delta}R^{\Delta}}{\Gamma(\Delta)R^{(d-1)/2}\ell_s^{\Delta}}\int_0^{\infty}\frac{d\beta}{\beta}\beta^{\Delta}e^{-2i\beta P^{\parallel}\cdot X^{\parallel}/\ell_s}e^{-2i\beta P^{\perp}\cdot X^{\perp}/\ell_s}\;,
\end{equation}
where we have separated the exponentials into a parallel part and a transverse part. Explicitly, the two components are defined by taking the first $p+2$ or the last $d-p$ elements of the embedding vectors 
\begin{equation}
    \begin{split}
        {}&X^\parallel=\frac{R}{z_0}\left(\frac{1+z_0^2+(\vec{z}^\parallel)^2+(\vec{z}^{\perp})^2}{2},\frac{1-z_0^2-(\vec{z}^\parallel)^2-(\vec{z}^{\perp})^2}{2},\vec{z}^\parallel,\vec{0}\right)\;,\quad  X^\perp=\frac{R}{z_0}\left(0,0,\vec{0},\vec{z}^\perp\right)\;,\\
        {}& P^\parallel=\left(\frac{1+(\vec{x}^\parallel)^2+(\vec{x}^\perp)^2}{2},\frac{1-(\vec{x}^\parallel)^2-(\vec{x}^\perp)^2}{2},\vec{x}^\parallel,\vec{0}\right)\;,\quad P^\perp=\left(0,0,\vec{0},\vec{x}^\perp\right)\;,
    \end{split}
\end{equation}
where we have used the Poincar\'e coordinates to parameterize the AdS point. The Minkowski space integrals over $X_1^\perp$, $Y^\parallel$, $Y^\perp$ amount to performing a Fourier transform which takes us into the momentum space and gives  
\begin{equation}
	\begin{split}
		I_{\mbb{M}}&=\int_{\mbb{M}}dX_1^{\perp}dY^{\parallel}dY^{\perp}e^{-i(2\beta_1 P_1^{\perp}\cdot X_1^{\perp}/\ell_s+2\beta_2 P_2^{\parallel}\cdot Y^{\parallel}/\ell_s+2\beta_2 P_2^{\perp}\cdot Y^{\perp}/\ell_s)}G_{\text{flat}}(X_1^\perp,Y^{\parallel},Y^{\perp})\\
		&=F_{\text{flat}}(k_1^{\parallel}=-k_2^{\parallel}=2\beta_2 P_2^{\parallel}/\ell_s,k_1^{\perp}=-2\beta_1P_1^{\perp}/\ell_s,k_2^{\perp}=-2\beta_2 P_2^{\perp}/\ell_s)\;.
	\end{split}
\end{equation}
Note that at the moment only $k_2$ is on-shell. The other momentum $k_1$ will become on-shell as well when we take the flat-space limit, through a similar mechanism as we have seen in the case without defects. We then consider the $X_1^\parallel$ integral over AdS$_{p+1}$
\begin{equation}\label{eq:IX_1}
		I_{X_1^{\parallel}}=\left(\prod_{i=1}^2\frac{(-i)^{\Delta_i}R^{\Delta_i}}{\Gamma(\Delta_i)R^{(d-1)/2}\ell_s^{\Delta_i}}\right)\int_{\te{AdS}_{p+1}}dW\int_0^{\infty}\prod_{i=1}^2\frac{d\beta_i}{\beta_i}\beta_i^{\Delta_i}e^{-2i\sum_{i=1}^2\beta_i P_i^{\parallel}\cdot W/\ell_s}\;,
\end{equation}
where we have renamed $X_1^\parallel$ as $W$ for notation simplicity. The $p$ integrals along the conformal boundary of the defect are Gaussian and can be easily performed to give
\begin{equation}\label{eq:integralforwa}
	\begin{split}
		I_{w^a}&=\int d^pw^a\exp\Big(i\beta_1R\frac{w_0^2+(x_1^i)^2+(x_1^a-w^a)^2}{w_0\ell_s}\Big)\exp\Big(i\beta_2R\frac{w_0^2+(x_2^i)^2+(x_2^a-w^a)^2}{w_0\ell_s}\Big)\\
		&=R^{p}\Big(\frac{i\pi w_0\ell_s}{R(\beta_1+\beta_2)}\Big)^{\frac{p}{2}}\exp\Big[i(\beta_1+\beta_2)w_0\frac{R}{\ell_s}\Big]\exp\Big[\frac{i\beta_1^2}{\beta_1+\beta_2}\frac{(x_1^2)^i}{w_0}\frac{R}{\ell_s}\Big]\exp\Big[\frac{i\beta_2^2}{\beta_1+\beta_2}\frac{(x_2^i)^2}{w_0}\frac{R}{\ell_s}\Big]\\
		&~~~\times\exp\Bigg[\frac{iR}{w_0\ell_s}\frac{\beta_1\beta_2}{\beta_1+\beta_2}\Big[(x_1^a-x_2^a)^2+(x_1^i)^2+(x_2^i)^2\Big]\Bigg]\;.
	\end{split}
\end{equation}
Note that here we have implicitly assumed $p\geq 1$ but the result in fact also applies to defects with $p=0$. In the $p=0$ case, the defect is created by a pair of heavy local operators in the CFT which we can choose to insert at $0$ and $\infty$. There are no integrals to be performed along the  boundary directions but the relevant part of the integrand in the two-point function\footnote{The integration along the defect still contains an integral in this case which is in the radial direction. The holographic dual of the $p=0$ defect is a geodesic line in AdS$_{d+1}$ which we can think of as a special limiting case of AdS$_1$.} turns out to be the same as in (\ref{eq:integralforwa}) by setting $p=0$. Let us introduce a convenient new variable $y=(\beta_1+\beta_2)w_0R/\ell_s$. Then the $X_1^\parallel$ integral becomes
\begin{equation}\label{IX1parallel}
	\begin{split}
		I_{X_1^{\parallel}}&=R^{p}\pi^{\frac{p}{2}}\left(\prod_{i=1}^2\frac{(-i)^{\Delta_i}R^{\Delta_i}}{\Gamma(\Delta_i)R^{(d-1)/2}\ell_s^{\Delta_i}}\right)\int_0^{\infty}\frac{d\beta_1}{\beta_1}\beta_1^{\Delta_1}\frac{d\beta_2}{\beta_2}\beta_2^{\Delta_2}\int_0^{\infty}\frac{dy}{y}(-iy)^{-\frac{p}{2}}e^{iy}\\
		&~~~\times\exp\Big[\frac{iR^2}{y\ell_s^2}\left(\beta_1^2(x_1^i)^2+\beta_2^2(x_2^i)^2+\left((x_1^a-x_2^a)^2+(x_1^i)^2+(x_2^i)^2\right)\beta_1\beta_2\right)\Big]\;.
	\end{split}
\end{equation}

\begin{figure}
	\centering
	\begin{tikzpicture}		
		\draw[dashed, gray] (0,-1.8) ellipse (2 and 0.4);
		\draw[dashed, gray] (0,1.8) ellipse (2 and 0.4);
		\draw[dashed, gray] (-2,1.8) -- (0,0) -- (2,-1.8);  % Lower cone
		\draw[dashed, gray] (-2,-1.8) -- (0,0) -- (2,1.8);    % Upper cone
		\filldraw[black] (-2,-1.8) circle (2pt) node[below] {$P_1$};
		\filldraw[black] (2,-1.8) circle (2pt) node[below] {$\bar{P}_1$};
		\filldraw[black] (1,2.15) circle (2pt) node[below] {$P_2$};
		\filldraw[black] (-0.8,1.43) circle (2pt) node[below] {$\bar{P}_2$};
		\draw[-,ultra thick] (0,-2)--(0,2);
	\end{tikzpicture}
	\caption{Configuration corresponding to the flat-space limit. Time goes along the vertical direction in the picture and the time-like defect is represented as the thick black line (with spatial directions omitted). This configuration allows us to find a point on the defect which is light-like separated from all operator insertions and their mirror images. 
	}
	\label{fig:configuration}
\end{figure}
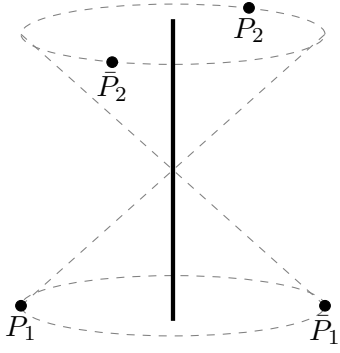

As in the defect-free case, the strategy for performing the $\beta_1$, $\beta_2$ integrals is to use the saddle point approximation which is justified by the large $R^2/\ell_s^2$ factor. However, we only get an enhanced behavior when one of the directions in the exponential is nearly flat. This requires the two operators to be inserted in a special configuration with respect to the defect and only this limit of the correlator will correspond to the flat-space limit in the bulk. To study this configuration in detail, let us define the following real symmetric matrix
\begin{equation}\label{matrixL}
	L=\begin{pmatrix} 
		-\frac{1}{2}P_1\cdot\bar{P}_1 & -\frac{1}{2}(P_1\cdot P_2+P_1\cdot\bar{P}_2)\\ 
		-\frac{1}{2}(P_1\cdot P_2+P_1\cdot\bar{P}_2) & -\frac{1}{2}P_2\cdot\bar{P}_2
	\end{pmatrix}\;.
\end{equation}
Here for each embedding space vector $P$ we have defined a mirror vector
\begin{equation}
	\bar{P}=\Big(\frac{1+(x^{\parallel})^2+(x^{\perp})^2}{2},\frac{1-(x^{\parallel})^2-(x^{\perp})^2}{2},x^{\parallel},-x^{\perp}\Big)\;,
\end{equation}
where the components transverse to the defect are reflected. Then in terms of this matrix the exponent can be compactly written as $\frac{iR^2}{y\ell_s^2}\sum_{m,n=1,2}L_{mn}\beta_m\beta_n$. For (\ref{matrixL}) to have a zero eigenvalue, a necessary condition is that the determinant of $L$ vanishes
\begin{equation}
    {\rm det}\,L=\frac{1}{4}\left((P_1\cdot\bar{P}_1)(P_2\cdot\bar{P}_2)-(P_1\cdot P_2+P_1\cdot\bar{P}_2)^2\right)=0\;.
\end{equation}
To understand the geometric meaning of this condition, let us first return to the defect-free case where the corresponding condition for matrix $P_{ij}$ takes the similar form of ${\rm det}\, P_{ij}=0$.
This condition can be analyzed by considering the following vector which is a linear combination of the four embedding vectors $P_i$
\begin{equation}
    Z=\sum_{i=1}^4 a_i P_i\;.
\end{equation}
Clearly, the condition of vanishing determinant is equivalent to requiring $Z$ to be orthogonal to all vectors $P_i$
\begin{equation}
    -2 Z\cdot P_i=\sum_{j=1}^4 a_j P_{ij}=0\;.
\end{equation}
The linear combination coefficients $a_i$ satisfying this condition form a null vector of the matrix $P_{ij}$. Moreover, it is easy to check that $Z$ obeys $Z\cdot Z=0$ and therefore is also a null embedding vector. Therefore, to take the flat-space limit the insertion points $P_i$ must be positioned in a configuration where they are all null separated from a common point in $\mathbb{R}^{1,d-1}$. Let us now turn to the defect case. We can consider a vector of the form
\begin{equation}
    Z=a_1(P_1+\bar{P}_1)+a_2(P_2+\bar{P}_2)\;.
\end{equation}
Similarly, we find that requiring the determinant of $L$ to vanish is the same as imposing the condition of $Z$ being orthogonal to all $P_i$ and $\bar{P}_i$
\begin{equation}
    \begin{split}
        {}&-\frac{1}{2}P_1\cdot Z=-\frac{1}{2}\bar{P}_1\cdot Z=L_{11}a_1+L_{12}a_2=0\;,\\ {}&-\frac{1}{2}P_2\cdot Z=-\frac{1}{2}\bar{P}_2\cdot Z=L_{21}a_1+L_{22}a_2=0\;.
    \end{split}
\end{equation}
When these relations are satisfied, $(a_1,a_2)$ is a null vector of the matrix $L$. It is also not difficult to verify that $Z$ is a null vector and therefore represents a point in $\mathbb{R}^{1,d-1}$. Note that by construction $Z$ has no transverse components which means it lies on the defect. Therefore, the flat-space limit configuration corresponds to having a point on the defect which is null separated from both $P_1$, $P_2$ and their mirrors $\bar{P}_1$, $\bar{P}_2$ (see Figure \ref{fig:configuration}). 

Similar to (\ref{configPi}) in the defect-free case, it is also convenient to work with an explicit configuration of $P_i$ which realizes the flat-space limit condition. For the moment, let us assume that the defect dimension satisfies $0\leq p<d-1$. The co-dimension 1 case, i.e., $p=d-1$,  is a bit special and will be discussed separately towards the end of this subsection. One convenient choice of the embedding vectors is
\begin{equation}\label{eq:flatspaceconfiguration}
	\begin{split}
		P_1&=(0,-1,-1,0,\vec{0})+\cdots\;,\\
		P_2&=(0,1,\cos\theta,\sin\theta,\vec{0})+\cdots\;,\\
		\bar{P}_1&=(0,-1,1,0,\vec{0})+\cdots\;,\\
		\bar{P}_2&=(0,1,-\cos\theta,-\sin\theta,\vec{0})+\cdots\;.
	\end{split}
\end{equation}
Here we have rotated the defect so that all of its space-like directions lie within the last $d-2$ components of the embedding vector. Therefore, the configuration appears to be identical to (\ref{configPi}) upon identifying $(P_1,P_2,P_3,P_4)$ in the defect-free case with $(P_1,\bar{P}_1,P_2,\bar{P}_2)$ in the defect case.
Using this parameterization, it is straightforward to find that the matrix $L$ becomes 
\begin{equation}
	L=\begin{pmatrix} 
		1 & -1\\ 
		-1 & 1
	\end{pmatrix}\;,
\end{equation}
and the exponent in (\ref{IX1parallel}) is diagonalized by switching to the new Schwinger parameters
\begin{equation}
        u=\frac{1}{\sqrt{2}}(\beta_1+\beta_2)\;,\quad \tilde{u}=\frac{1}{\sqrt{2}}(-\beta_1+\beta_2)\;.
\end{equation}
These two combinations correspondingly have eigenvalues
\begin{equation}\label{eigenvalueL}
    \lambda=0+\ldots\;,\quad \tilde{\lambda}=2+\ldots\;.
\end{equation}

After this small detour, let us now resume the analysis of the integral where we left at (\ref{IX1parallel}). As in the defect-free case, we need to analyze the small vicinity of this point so that $\lambda$ is small but nonzero. In the new variables, the $\beta_i$ integrals become
\begin{equation}
\begin{split}
	I_{\beta}={}&2^{1-\frac{\Delta_1+\Delta_2}{2}}\int_0^{\infty} du\int_{-u}^ud\tilde{u}(u-\tilde{u})^{\Delta_1-1}(u+\tilde{u})^{\Delta_2-1}e^{i\frac{R^2}{y\ell_s^2}(\lambda u^2+\tilde{\lambda}\tilde{u}^2)}\\
	={}&2^{-\frac{-1+\Delta_1+\Delta_2}{2}}\frac{i\ell_s\sqrt{\pi}}{R}(-i y)^{\frac{1}{2}}\int_0^{\infty}u^{\Delta_1+\Delta_2-2}e^{i\frac{R^2}{y\ell_s^2}\lambda u^2}\;,
\end{split}
\end{equation}
where in the second step we have evaluated the $\tilde{u}$ integral by saddle point. The two-point function (\ref{defF2pt}) can then be written as
\begin{equation}\label{eq:Dzetasigmayu}
	\begin{split}
		G(\eta,\theta)&=2^{-\frac{-1+\Delta_1+\Delta_2}{2}}\pi^{\frac{p+1}{2}}R^{p+2-d}\left(\frac{R}{\ell_s}\right)^{\Delta_1+\Delta_2-1}\frac{(-i)^{\Delta_1+\Delta_2-1}}{\Gamma(\Delta_1)\Gamma(\Delta_2)}\\
		&~~~\times\int_0^{\infty}du u^{\Delta_1+\Delta_2-2}\int_0^{\infty}\frac{dy}{y}(-iy)^{\frac{1-p}{2}}e^{iy}e^{-i\frac{\eta^2}{4y}u^2}F_{\te{flat}}(S,Q)\;.
	\end{split}	
\end{equation}
Here we have defined the combination 
\begin{equation}\label{defetaflambda}
    \eta^2=-4\frac{R^2}{\ell_s^2}\lambda\;,
\end{equation}
which is kept finite in taking the flat-space limit. Note that in this limit $k_1$ also becomes on-shell
\begin{equation}
    k_1^2=\beta_1^2 (P_1-\bar{P}_1)^2+\beta_2^2(P_2+\bar{P}_2)^2=4\beta_1^2-4\beta_2^2=-8u\tilde{u}\approx 0\;.
\end{equation}
This allows us to write the flat-space form factor in (\ref{eq:Dzetasigmayu}) in terms of the defect Mandelstam variables 
\begin{equation}\label{eq:QSspecial}
	Q=-(\vec{k}_1^{\parallel})^2=-(\vec{k}_2^{\parallel})^2=\frac{2u^2}{\ell_s^2} \;,\quad\quad S=-\vec{k}_1\cdot\vec{k}_2=-\frac{2u^2}{{\ell_s^2}}(1-\cos\theta)\;.
\end{equation}
Finally, we can perform the integral over $y$ and obtain the following relation between the defect two-point function and the flat-space form factor
\begin{equation}\label{eq:Datflatspace}
	\begin{split}
		G(\eta,\theta)=&\left(\frac{R}{\ell_s}\right)^{\Delta_1+\Delta_2+p+1-d}\frac{(-i)^{\Delta_1+\Delta_2-1}\pi^{\frac{p+1}{2}}}{\Gamma(\Delta_1)\Gamma(\Delta_2)2^{\frac{\Delta_1+\Delta_2-p-2}{2}}}\\
		&~~~\times\int_0^{\infty}du u^{\Delta_1+\Delta_2-2}(u\eta)^{\frac{1-p}{2}}K_{\frac{p-1}{2}}(u\eta)\ma{F}_{\te{flat}}\left(S=-\frac{2u^2}{\ell_s^2}(1-\cos\theta),Q=\frac{2u^2}{\ell_s^2}\right)\;.
	\end{split}
\end{equation}
Here we have extracted a power of $\ell_s$ to make the flat-space form factor dimensionless
\begin{equation}
    F_{\te{flat}}=\ell_s^{d-p-2}\ma{F}_{\te{flat}}\;.
\end{equation}

It is important to point out that although we derived this formula using the explicit parameterization (\ref{eq:flatspaceconfiguration}), the result applies generally to all conformally equivalent flat-space limit configurations. To state this more precisely, let us write the defect two-point function in terms of the defect conformal cross ratios (see \cite{Billo:2016cpy} for a more detailed discussion of the conformal kinematics of defect correlators)
\begin{equation}\label{defcalG}
    G(P_1,P_2)=\frac{1}{(-\frac{1}{2}P_1\cdot \bar{P}_1)^{\frac{\Delta_1}{2}}(-\frac{1}{2}P_2\cdot \bar{P}_2)^{\frac{\Delta_2}{2}}}\mathcal{G}(\xi,\chi)\;,
\end{equation}
where
\begin{equation}
    \xi=\frac{-4P_1\cdot P_2}{\sqrt{(P_1\cdot \bar{P_1})(P_2\cdot \bar{P}_2)}}=\frac{x_{12}^2}{|x_1^i||x_2^i|}\;,\quad \chi=\frac{2(P_1\cdot P_2-P_1\cdot \bar{P}_2)}{\sqrt{(P_1\cdot \bar{P_1})(P_2\cdot \bar{P}_2)}}=\frac{2x_1^jx_2^j}{|x_1^i||x_2^i|}\;.
\end{equation}
To make the connection with the special configuration more clear, we note that we can use conformal symmetry to perform a rescaling in (\ref{IX1parallel})
\begin{equation}
    \beta_1\to \frac{\beta_1}{\sqrt{(-\frac{1}{2}P_1\cdot \bar{P}_1)}}\;,\quad   \beta_2\to \frac{\beta_2}{\sqrt{(-\frac{1}{2}P_2\cdot \bar{P}_2)}}\;.
\end{equation}
This makes the matrix $L$ invariant with respect to the independent rescalings of the embedding space vectors $P_i \to \Lambda_i P_i$, $\bar{P}_i\to \Lambda_i \bar{P}_i$
\begin{equation}
	L\to L'=\begin{pmatrix} 
		1 & \frac{-\frac{1}{2}(P_1\cdot P_2+P_1\cdot\bar{P}_2)}{\sqrt{(-\frac{1}{2}P_1\cdot \bar{P}_1)(-\frac{1}{2}P_2\cdot \bar{P}_2)}}\\ 
		\frac{-\frac{1}{2}(P_1\cdot P_2+P_1\cdot\bar{P}_2)}{\sqrt{(-\frac{1}{2}P_1\cdot \bar{P}_1)(-\frac{1}{2}P_2\cdot \bar{P}_2)}} & 1
	\end{pmatrix}\;,
\end{equation}
and also generates the one-point function factors which are extracted in (\ref{defcalG}). Note that the rescaling factors are trivial for the configuration (\ref{eq:flatspaceconfiguration}) and therefore do not change the eigenvalues (\ref{eigenvalueL}). To proceed, we can make a convenient change of variable to replace $\xi$ in $\mathcal{G}$ by
\begin{equation}
    \Theta={\rm det} \,L'=1-\frac{(\xi+\chi)^2}{4}\;.
\end{equation}
In the flat-space limt,  we need to take the limit $\Theta\to 0$ but keep fixed 
\begin{equation}
    \eta^2=-2\frac{R^2}{\ell_s^2}\Theta\;,
\end{equation}
where $\eta$ should be identified the same variable introduced in (\ref{defetaflambda}) for the special configuration. We also have 
\begin{equation}
    \chi=-2\left(1+\frac{S}{Q}\right)\;,
\end{equation}
which expresses the ratio of flat-space Mandelstam variables in terms of the conformal cross ratio. The scaling limit of the correlator relevant for the flat-space limit can be more precisely defined as
\begin{equation}
    \mathcal{G}_{\rm f.s.}(\eta,\chi)=\lim_{R/\ell_s\to\infty}\left(\frac{R}{\ell_s}\right)^{-(\Delta_1+\Delta_2+p+1-d)}\mathcal{G}\left(\Theta=-\frac{\eta^2\ell_s^2}{2R^2},\chi\right)\;.
\end{equation}
Our result (\ref{eq:Datflatspace}) then gives the position space flat-space limit formula which relates the defect two-point correlator in this scaling limit with the flat-space form factor
\begin{equation}\label{defectfslimit}
\begin{split}
    \mathcal{G}_{\rm f.s.}(\eta,\chi)=&\frac{(-i)^{\Delta_1+\Delta_2-1}\pi^{\frac{p+1}{2}}}{\Gamma(\Delta_1)\Gamma(\Delta_2)2^{\frac{\Delta_1+\Delta_2-p-2}{2}}}\int_0^{\infty}du u^{\Delta_1+\Delta_2-2}(u\eta)^{\frac{1-p}{2}}\\
    &\times K_{\frac{p-1}{2}}(u\eta)\ma{F}_{\te{flat}}\left(S=-\frac{2+\chi}{\ell_s^2}u^2,Q=\frac{2u^2}{\ell_s^2}\right)\;.
\end{split}
\end{equation}

In the derivation, we have assumed $0\leq p\leq d-2$. Let us now comment on the special case with $p=d-1$ which corresponds to a boundary or an interface. In this case, there is only one transverse component which allows us to replace $|x_1^i||x_2^i|$ with $|x_1^ix_2^i|$. This eliminates one of the conformal cross ratios by setting it to a constant value
\begin{equation}
    \chi=2\;.
\end{equation}
The condition of vanishing determinant can be rewritten simply as 
\begin{equation}
    P_1\cdot \bar{P}_2=0\;.
\end{equation}
The flat-space limit therefore requires one operator to approach the mirror image of the other operator. This limit was also studied in \cite{Mazac:2018biw} where it was termed the Regge limit and its importance was recognized in the context of the analytic functional method. On the other hand, we note that in flat space we can solve $k_i^\perp$ in terms of $k_i^\parallel$ from the on-shell condition. This implies that only one Mandelstam variable is independent, i.e., $2Q=-S$. We can redo the analysis of the flat-space limit with the additional kinematic constraint of co-dimension 1 defects taken into account. However, we find that the result is identical to (\ref{defectfslimit}) except that we need to keep only one conformal cross ratio and one Mandalstam variable.  

For completeness, let us also record the inverse transformation of (\ref{defectfslimit}). To this end, we can use the following orthogonality relation of the modified Bessel function\footnote{It is important to note that the orthogonality relation does not hold pointwise, but rather in the distributional sense. The contour runs from $-i\infty$ to $i \infty$ and the small $c>0$ is to avoid the singularity at $\eta=0$. This identity can be proven by multiplying the differential equation
\begin{equation}
\frac{1}{\eta}\frac{d}{d\eta}\left(\eta\frac{d K_\nu(u\eta)}{d\eta}\right)-\left(u^2+\frac{\nu^2}{\eta^2}\right)K_\nu(u\eta)=0\;.
\end{equation}
with $\eta K_\nu(-v\eta)$ and then integrating by parts.}
\begin{equation}
	\int_{c-i\infty}^{c+i\infty}d\eta \eta K_{\nu}(u\eta)K_{\nu}(-v\eta)=\frac{\pi^2}{u}\delta(u-v)\;,
\end{equation}
where $u>0$ and $v>0$. This gives
\begin{equation}\label{eq:flatformulaF}
		\ma{F}_{\te{flat}}(S,Q)=\frac{i^{\Delta_1+\Delta_2-1}\Gamma(\Delta_1)\Gamma(\Delta_2)}{\pi^{\frac{p+5}{2}}2^{\frac{p-\Delta_1-\Delta_2+2}{2}}}u^{\frac{5+p}{2}-\Delta_1-\Delta_2} \int_{c-i\infty}^{c+i\infty}d\eta \eta^{\frac{p+1}{2}}K_{\frac{p-1}{2}}(-u\eta)\ma{G}_{\te{f.s.}}(\eta,\chi)\;,
\end{equation}
with 
\begin{equation}
    u^2=\frac{Q\ell_s^2}{2}\;,\quad S=-\frac{2+\chi}{\ell_s^2}u^2\;.
\end{equation}

\subsection{Analytic continuation}\label{Sec:analconti}
It is clear that the flat-space limit can only be taken when the spacetime is Lorentzian. However, holographic correlators are usually more convenient to compute in the Euclidean signature. Therefore, a nontrivial analytic continuation is needed to go from Euclidean results to the Lorentzian ones in order to obtain  flat-space limit amplitudes from holographic correlators. In this subsection, we will analyze this problem in detail and give the prescription for defect two-point functions. 

We can proceed in the same way as in four-point functions in defect-free CFTs. In global coordinates, the path of the Wick rotation takes the form \cite{Gary:2009ae}
\begin{equation}
    P\to \left(\cos(-i\tau e^{i\alpha}),\sin(-i\tau e^{i\alpha}),{\bf e}\right)\;,
\end{equation}
where the global time $\tau$ is rotated as $\tau\to -i\tau e^{i\alpha}$. The analytic continuation parameter $\alpha$ takes value in the range $[0,\frac{\pi}{2}]$ where $\alpha=0$ is Euclidean and $\alpha=\frac{\pi}{2}$ is Lorentzian.\footnote{Note that the signature of the space is kept unchanged and is $(-,-,+,\ldots,+)$.} We will focus on two sets of variables and observe how they are analytically continued as we change $\alpha$. 

\begin{figure}[htbp]
\begin{center}
\includegraphics[width=0.5\textwidth]{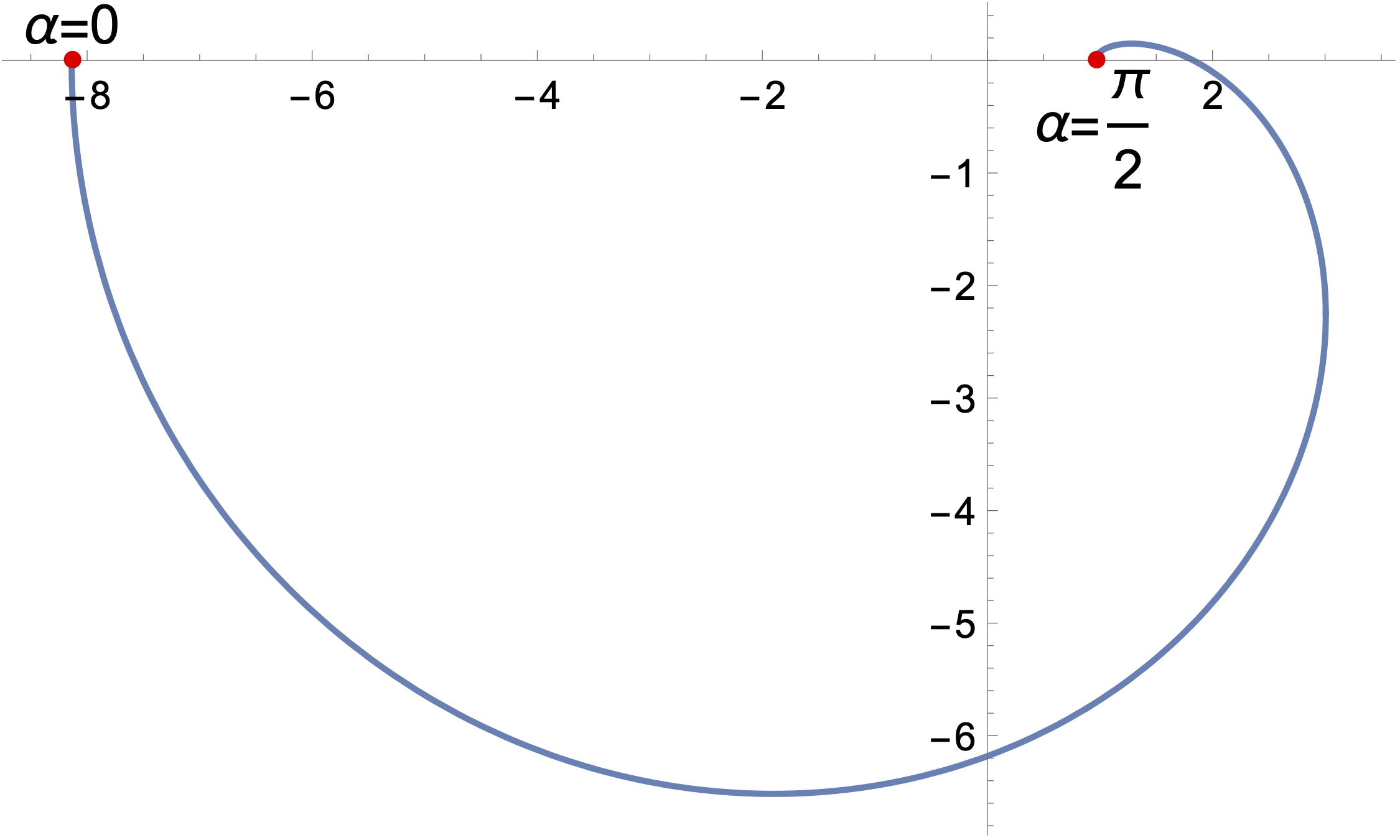}
\caption{Analytic continuation of $\Xi$. Here we have set $\epsilon=0.2$.}
\label{fig_pathXi}
\end{center}
\end{figure}

The first variable of interest is the following combination
\begin{equation}
    \Xi=-\frac{\xi+\chi-2}{4}\;.
\end{equation}
This combination appears in the flat-space limit discussed in the previous subsection where it approaches the value $1$. It also appears as the argument in the contact Witten diagram \cite{Rastelli:2017ecj,Gimenez-Grau:2023fcy}
\begin{equation}\label{Gcontact}
    \ma{G}^{\rm contact}_{\Delta_1\Delta_2}=\left(\frac{R}{\ell_s}\right)^{p-d+2}\frac{\pi^{\frac{p+1}{2}}\Gamma(\frac{\Delta_1+\Delta_2-p}{2})}{2^{\Delta_1+\Delta_2}\Gamma(\frac{\Delta_1+\Delta_2+1}{2})}{}_2F_1\left(\Delta_1,\Delta_2,\frac{\Delta_1+\Delta_2+1}{2},\Xi\right)\;.
\end{equation}
Note that in the Euclidean regime $\Xi$ is restricted to the region
\begin{equation}
    \Xi=-\frac{(x_1^a-x_2^a)^2+(|x_1^i|-|x_2^i|)^2}{4|x_1^i||x_2^i|}\in (-\infty,0]\;.
\end{equation}
To reach $\Xi=1$ an analytic continuation is needed. Let us consider the following configuration
\begin{equation}\label{eq:configanalcont}
	\begin{split}
		P_1&=\left(\cos(-i\tau_1 e^{i\alpha}),\sin(-i\tau_1 e^{i\alpha}),-1,0,\vec{0}\right),\\
		P_2&=\left(\cos(-i\tau_2 e^{i\alpha}),\sin(-i\tau_2 e^{i\alpha}),\cos\theta,\sin\theta,\vec{0}\right),\\
		\bar{P}_1&=\left(\cos(-i\tau_1 e^{i\alpha}),\sin(-i\tau_1 e^{i\alpha}),1,0,\vec{0}\right),\\
		\bar{P}_2&=\left(\cos(-i\tau_2 e^{i\alpha}),\sin(-i\tau_2 e^{i\alpha}),-\cos\theta,-\sin\theta,\vec{0}\right),
	\end{split}
\end{equation}
where  (\ref{eq:flatspaceconfiguration}) is the special case with $\tau_1=-\frac{\pi}{2}$, $\tau_2=\frac{\pi}{2}$ and $\alpha=\frac{\pi}{2}$. This gives
\begin{equation}
    \Xi=-\sinh^2\left(\frac{(\tau_1-\tau_2)e^{i\alpha}}{2}\right)\;.
\end{equation}
To correctly perform the analytic continuation, we should take $\tau_1=-\frac{\pi}{2}-\epsilon$ and $\tau_2=\frac{\pi}{2}+\epsilon$ for some small positive $\epsilon$. Then we follow the path of analytic continuation where $\alpha$ goes from $0$ to $\frac{\pi}{2}$ (Figure \ref{fig_pathXi}). We find that the trajectory crosses the branch cut $[1,\infty)$ from below and approaches again the real axis from above where it intersects to the left of $1$.

This analysis allows us to take the flat-space limit of contact Witten diagrams. After analytic continuation, we can reach the point $\Xi \to 1$ in (\ref{Gcontact}). By using the identity
\begin{equation}
\begin{split}
    {}_2F_1(a,b,c,z)={}&\frac{\Gamma(c)\Gamma(a+b-c)}{\Gamma(a)\Gamma(b)}(1-z)^{c-a-b}{}_2F_1(c-a,c-b,c-a-b+1,1-z)\\
    {}&+\frac{\Gamma(c)\Gamma(c-a-b)}{\Gamma(c-a)\Gamma(c-b)}{}_2F_1(a,b,a+b-c+1,1-z)\;,
\end{split}
\end{equation}
we find that the contact Witten diagram has a leading power singularity
\begin{equation}
    \mathcal{G}^{\rm contact}_{\Delta_1\Delta_2}\sim \left(\frac{R}{\ell_s}\right)^{p-d+2}\frac{\pi^{\frac{p+1}{2}}\Gamma(\frac{\Delta_1+\Delta_2-p}{2})\Gamma(\frac{\Delta_1+\Delta_2-1}{2})}{2^{\Delta_1+\Delta_2}\Gamma(\Delta_1)\Gamma(\Delta_2)}(1-\Xi)^{\frac{1-\Delta_1-\Delta_2}{2}}\;,
\end{equation}
while the subleading terms are suppressed by extra powers of $\ell_s/R$ when switching from $1-\Xi$ to $-\eta^2=8(\frac{R}{\ell_s})^2\Xi(1-\Xi)$. The flat-space limit formula (\ref{eq:flatformulaF}) then gives
\begin{equation}
    \mathcal{F}_{\rm flat, con}=1\;,
\end{equation}
as expected, where we have used the identity
\begin{equation}
\int_{c-i\infty}^{c+i\infty} d\eta K_\nu(-u\eta) \eta^{1-\alpha}=\pi^2\left(\frac{u}{2}\right)^{\alpha-2}\frac{1}{\Gamma(\frac{\alpha-\nu}{2})\Gamma(\frac{\alpha+\nu}{2})}\;.
\end{equation}

\begin{figure}[htbp]
\begin{center}
\includegraphics[width=0.5\textwidth]{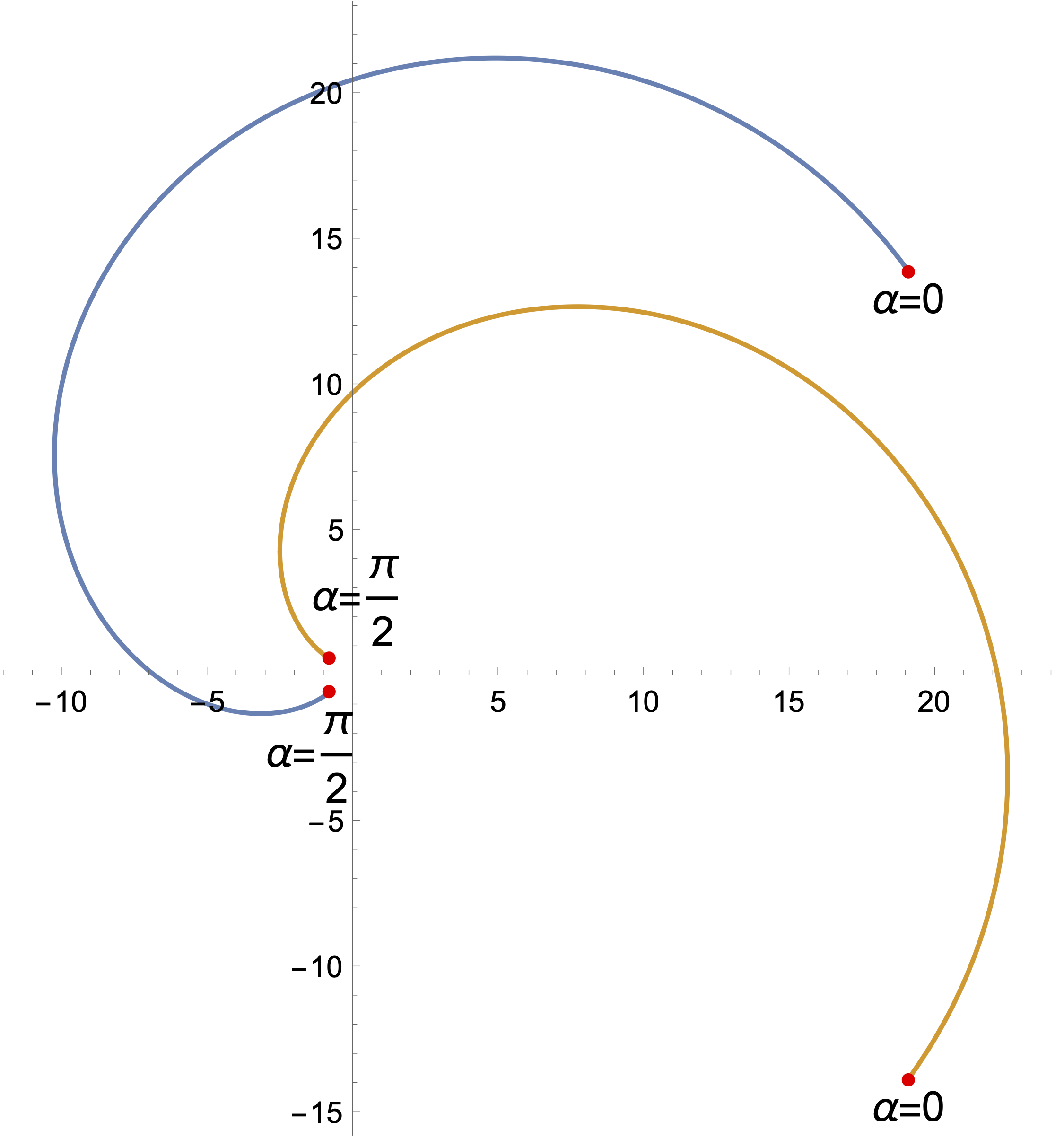}
\caption{Analytic continuation of $x$ (in orange) and $\bar{x}$ (in blue). Here we have set $\theta=\frac{\pi}{5}$ and $\epsilon=0.01$.}
\label{fig_pathzzb}
\end{center}
\end{figure}

Note that in general a defect two-point function depends on two cross ratios instead of one, while we have so far only considered one particular combination. To this end, let us consider the following set of variables $x$, $\bar{x}$ defined via 
\begin{equation}\label{defzzb}
    \xi+\chi=\frac{1+x\bar{x}}{\sqrt{x\bar{x}}}\;,\quad\quad \chi=\frac{x+\bar{x}}{\sqrt{x\bar{x}}}\;.
\end{equation}
The new cross ratios $x$, $\bar{x}$ are similar to the $z$, $\bar{z}$ cross ratios defined for defect-free four-point functions. In fact, we can make the connection more precise. Using conformal symmetry, we can put $x_1$ and $x_2$ on a 2d plane and require the defect to intersect the plane perpendicularly at $0$ and $\infty$.\footnote{This requires the defect to have at least co-dimension 2. However, when the co-dimension is 1, we have only one independent cross ratio and the discussion of $\Xi$ is  sufficient.} We can further move $x_1$ to be at $1$ and the complex coordinates of $x_2$ then become $(x,\bar{x})$. Therefore, the precise relations are
\begin{equation}\label{zzbandxxb}
    z=1-x\;,\quad \quad \bar{z}=1-\bar{x}\;,
\end{equation}
where the four-point function cross ratios are defined as
\begin{equation}
    z=\frac{(x_1-x_2)(x_3-x_4)}{(x_1-x_3)(x_2-x_4)}\;,\quad\quad \bar{z}=\frac{(\bar{x}_1-\bar{x}_2)(\bar{x}_3-\bar{x}_4)}{(\bar{x}_1-\bar{x}_3)(\bar{x}_2-\bar{x}_4)}\;.
\end{equation}
Note that $x\to 1$ is the bulk channel OPE limit and $x\to 0$ or $x\to \infty$ is the defect channel OPE limit. We also point out that this parameterization has a four-fold redundancy because $\xi$, $\chi$ are invariant under the action of the two generators
\begin{equation}
    (x,\bar{x})\to (\bar{x},x)\;,\quad (x,\bar{x})\to (1/x,1/\bar{x})\;.
\end{equation}
The latter generator has a simple physical interpretation. Under the conformal inversion $x\to 1/x$, the two intersecting points of the defect with the plane $0$ and $\infty$ are interchanged. However, this should not change the defect because these two points are indistinguishable. Note also $x=1$ is a fixed point of the inversion. Therefore, the invariance of the correlator requires $(x,\bar{x})\to (1/x,1/\bar{x})$ to be a redundancy of the parameterization. This redundancy allows us to choose equivalently any of the four solutions when we solve $\xi$, $\chi$ in terms of $x$, $\bar{x}$. For the configuration (\ref{eq:configanalcont}), we have 
\begin{equation}
    x=e^{i\theta+e^{i\alpha}(-\tau_1+\tau_2)}\;,\quad\quad  \bar{x}=e^{-i\theta+e^{i\alpha}(-\tau_1+\tau_2)}\;.
\end{equation}
The path of analytic continuation is illustrated in Figure \ref{fig_pathzzb}. We can see that going from Euclidean to Lorentzian, $x$ crosses only the branch cut $[1,\infty)$ of the defect channel conformal blocks while $\bar{x}$ crosses only the branch cut $(-\infty,0]$ of the bulk channel conformal blocks.
 
\subsection{Equivalence with the Mellin space formula}\label{Sec:equivwithMellin}
Mellin space \cite{Mack:2009mi,Penedones:2010ue} provides a convenient way to represent holographic correlators by simplying their analytic structure. This formalism was generalized in \cite{Rastelli:2017ecj,Goncalves:2018fwx} to include boundaries and defects. In \cite{Alday:2024srr} a flat-space limit formula was conjectured in Mellin space. In this subsection, we will prove the Mellin space formula by showing that it is equivalent to our position space formula.

Let us start with the definition of Mellin space for two-point functions
\begin{equation}\label{defMellinrep2pt}
    \mathcal{G}(\xi,\chi)=\frac{\pi^{\frac{p}{2}}\Gamma(\frac{\Delta_1+\Delta_2-p}{2})}{4\Gamma(\Delta_1)\Gamma(\Delta_2)}\int\frac{d\delta d\gamma}{(2\pi i)^2}\xi^{-\delta}\chi^{-\gamma+\delta}\ma{M}(\delta,\gamma)\Gamma(\delta)\Gamma(\gamma-\delta)\prod_{i=1}^2\Gamma\Big(\frac{\Delta_i-\gamma}{2}\Big)\;.
\end{equation}
We parameterize the contours as
\begin{equation}
    \delta=c+i \omega\;,\quad \gamma=2c+i\upsilon\;,
\end{equation}
where $c$ is a small positive number. For large $\omega$ and $\upsilon$, the Gamma factor provides an exponential suppression 
\begin{equation}
    \bigg|\Gamma(\delta)\Gamma(\gamma-\delta)\prod_{i=1}^2\Gamma\Big(\frac{\Delta_i-\gamma}{2}\Big)\bigg|\sim e^{-\frac{\pi}{2}(|\omega|+|\upsilon-\omega|+|\upsilon|)}\;.
\end{equation}
As long as the Mellin amplitude does not grow exponentially fast at infinities, this suppression ensures the convergence of the Mellin representation (\ref{defMellinrep2pt}). In the Euclidean regime where $\xi>0$ and $\chi>0$, the factor of cross ratios only gives rise to oscillations for large $\omega$ and $\upsilon$. However, when we analytically continue to Lorentzian regime this factor can provide an exponential growth. More precisely, to realize the flat-space limit we need to require $P_1$ and $P_2$ to be time-like separated which dictates $\xi<0$. On the other hand, the other cross ratio $\chi=-2\cos\theta_{12}$, with $\theta_{12}$ being the angle between the angle between 1 and 2 in the directions perpendicular to the defect, can take either sign. Therefore, we need to distinguish two cases and we have
\begin{equation}
    \xi^{-\delta}\chi^{-\gamma+\delta}\sim\left\{\begin{array}{l}e^{\pi \omega} \quad\quad \chi>0\;,\\ e^{\pi\upsilon}\quad\quad \chi<0\;,\end{array}\right.
\end{equation}
for large $\omega$ and $\upsilon$. Let us write $\upsilon=\kappa \omega$. Then it is easy to verify that the suppression is precisely cancelled when
\begin{equation}\label{kapparange}
    \begin{split}
        {}&\chi>0\;:\quad 0\leq\kappa\leq 1\;,\quad \omega>0\;,\\
        {}&\chi<0\;:\quad \kappa\geq 1\;,\quad \omega>0\;.
    \end{split}
\end{equation}
These are precisely the regimes where we get an enhancement for the flat-space limit. Let us analyze the cross raito factor and the Gamma function factor in more details. In both cases, they give
\begin{equation}
   \xi^{-\delta}\chi^{-\gamma+\delta} \Gamma(\delta)\Gamma(\gamma-\delta)\prod_{i=1}^2\Gamma\Big(\frac{\Delta_i-\gamma}{2}\Big)\approx \frac{(2\pi)^2\left(-\frac{i\kappa\omega}{2}\right)^{\frac{\Delta_1+\Delta_2-2}{2}}}{\sqrt{1-\kappa}\omega}e^{-i\omega(\log(\frac{(\kappa-1)\xi}{\chi})+\kappa\log(\frac{\kappa\chi}{2(1-\kappa)}))}\;,
\end{equation}
but $\kappa$ is in different ranges according to (\ref{kapparange}). For large $\omega$ we can integrate out $\kappa$ by using the saddle point approximation and the dominant contribution comes from the saddle point
\begin{equation}
    \kappa=\kappa_*=\frac{2}{\chi+2}\;.
\end{equation}
The $\kappa$ integral gives
\begin{equation}
		\ma{G}(\xi,\chi)\approx \frac{\pi^{\frac{p+1}{2}}\Gamma(\frac{\Delta_1+\Delta_2-p}{2})}{2\Gamma(\Delta_1)\Gamma(\Delta_2)}\int_0^{\infty}\frac{d\omega}{\omega}\Big(\frac{-i\omega}{2+\chi}\Big)^{\frac{\Delta_1+\Delta_2-1}{2}}e^{-i\omega\log\frac{-\xi}{2+\chi}}\ma{M}(i\omega,i\kappa_* \omega)\;.
\end{equation}
For general values of the cross ratios, the integral over $\omega$ is violently oscillating and the correlator is suppressed. However, in the flat-space limit where we need zoom in around the configuration 
\begin{equation}
    2+\chi=-\xi\;,
\end{equation}
which makes the logarithm vanish and enhances the integral. To analyze the integral around this point more precisely, let us introduce the following new variables
\begin{equation}
  \sqrt{x}=i \rho^{1/2}e^{\iota/2}\;,\quad\sqrt{\bar{x}}=i \rho^{-1/2}e^{\iota/2}\;,  
\end{equation}
where $x$, $\bar{x}$ are related to $\xi$, $\chi$ as in (\ref{defzzb}). The flat-space limit corresponds to taking $\iota\to 0$ and the correlator in this limit reduces to
\begin{equation}\label{omegaintegral}
		\ma{G}(\xi,\chi)\approx \frac{\pi^{\frac{p+1}{2}}\Gamma(\frac{\Delta_1+\Delta_2-p}{2})}{2\Gamma(\Delta_1)\Gamma(\Delta_2)}\int_0^\infty\frac{d\omega}{\omega}\Big(\frac{-i\omega}{2+\chi}\Big)^{\frac{\Delta_1+\Delta_2-1}{2}}e^{-\frac{i}{2}\omega\iota^2\kappa_*}\ma{M}(i\omega,i\kappa_* \omega)\;.
\end{equation}
On the other hand, it was conjectured in \cite{Alday:2024srr} that  the flat-space form factor is related to the high energy limit of the Mellin amplitude by 
\begin{equation}
	\ma{M}(\delta,\gamma)=\frac{(R/\ell_s)^{p-d+2}}{\Gamma(\frac{\Delta_1+\Delta_2-p}{2})}\int_0^{\infty}\frac{d\beta}{\beta}\beta^{\frac{\Delta_1+\Delta_2-p}{2}}e^{-\beta}\ma{F}_{\te{flat}}\Big(S=-\frac{2\delta\beta}{R^2},Q=\frac{2\gamma\beta}{R^2}\Big)\;,
\end{equation}
where $R/\ell_s\to\infty$ and $S$, $Q$ are kept fixed. Substituting it into (\ref{omegaintegral}), we get 
\begin{equation}
	\begin{split}
		\ma{G}(\xi,\chi)\approx& \frac{\pi^{\frac{p+1}{2}}}{2\Gamma(\Delta_1)\Gamma(\Delta_2)}\int_0^\infty\frac{d\omega}{\omega}\Big(\frac{-i\omega}{2+\chi}\Big)^{\frac{\Delta_1+\Delta_2-1}{2}}e^{-\frac{i}{2}\omega\iota^2\kappa_*}\\
		&\times \left(\frac{R}{\ell_s}\right)^{p-d+2}\int_0^{\infty}\frac{d\beta}{\beta}\beta^{\frac{\Delta_1+\Delta_2-p}{2}}e^{-\beta}\ma{F}_{\te{flat}}\Big(\frac{Q}{S}=-\kappa_*,Q=\frac{2i\kappa_*\omega\beta}{R^2}\Big)\;.
	\end{split}
\end{equation}
We rescale $\omega\to \frac{\omega R^2}{\ell_s^2\kappa_*}$ and then $\beta\to\beta/\omega$. This gives
\begin{equation}
	\begin{split}
		\ma{G}(\xi,\chi)\approx& \frac{\pi^{\frac{p+1}{2}}}{2\Gamma(\Delta_1)\Gamma(\Delta_2)}\left(\frac{-iR^2}{2\ell_s^2}\right)^{\frac{\Delta_1+\Delta_2-1}{2}}\int_0^\infty\frac{d\omega}{\omega}\omega^{\frac{p-1}{2}}e^{-\frac{i}{4}\omega\eta^2}\\
		&\times \left(\frac{R}{\ell_s}\right)^{p-d+2}\int_0^{\infty}\frac{d\beta}{\beta}\beta^{\frac{\Delta_1+\Delta_2-p}{2}}e^{-\frac{\beta}{\omega}}\ma{F}_{\te{flat}}\Big(\frac{Q}{S}=-\kappa_*,Q=\frac{2i\beta}{\ell_s^2}\Big)\;,
	\end{split}
\end{equation}
where we have used $\eta^2=2\iota^2\frac{R^2}{\ell_s^2}$. The integration over $\omega$ gives rise to a Bessel $K$ function
\begin{equation}
	\begin{split}
		\ma{G}(\xi,\chi)\approx& \frac{\pi^{\frac{p+1}{2}}(-i)^{\Delta_1+\Delta_2-1}}{2\Gamma(\Delta_1)\Gamma(\Delta_2)2^{\frac{\Delta_1+\Delta_2-p-2}{2}}}\left(\frac{R}{\ell_s}\right)^{\Delta_1+\Delta_2+1+p-d}\\
		&\times \int_0^{\infty}\frac{d\beta}{\beta}(i\beta)^{\frac{\Delta_1+\Delta_2-1}{2}-\frac{p-1}{4}}K_{\frac{p-1}{2}}(\eta\sqrt{i\beta})\eta^{\frac{1-p}{2}}\ma{F}_{\te{flat}}\Big(\frac{Q}{S}=-\kappa_*,Q=\frac{2i\beta}{\ell_s^2}\Big)\;.
	\end{split}
\end{equation}
We Wick rotate the contour from $\beta\in \mathbb{R}^+$ to $i\beta\in \mathbb{R}^+$ and make the change of variable $u^2=i\beta$. This gives
\begin{equation}
\begin{split}
    \mathcal{G}(\xi,\chi)\approx{}&\frac{\pi^{\frac{p+1}{2}}(-i)^{\Delta_1+\Delta_2-1}}{\Gamma(\Delta_1)\Gamma(\Delta_2)2^{\frac{\Delta_1+\Delta_2-p-2}{2}}}\left(\frac{R}{\ell_s}\right)^{\Delta_1+\Delta_2+1+p-d}\\
    {}&\times\int_0^\infty du u^{\Delta_1+\Delta_2-2} (u\eta)^{\frac{1-p}{2}}K_{\frac{p-1}{2}}(u\eta)\mathcal{F}_{\rm flat}\Big(\frac{Q}{S}=-\frac{2}{\chi+2},Q=\frac{2u^2}{\ell_s^2}\Big)\;,
\end{split}
\end{equation}
which reproduces (\ref{defectfslimit}).

\section{Explicit examples}\label{Sec:explicitexamples}
In this section we check our position space formula in a few nontrivial examples. These examples are from top-down models of holographic defects in string theory and M-theory, and include Wilson loops (or 't Hooft loops) \cite{Gimenez-Grau:2023fcy} and giant gravitons in 4d $\mathcal{N}=4$ SYM \cite{Chen:2025yxg}, and surface defects in the 6d $(2,0)$ theory \cite{Chen:2023yvw}. In these setups, the defects have dimensions 1, 0 and 2 respectively. The flat-space limits of the line defect case and surface defect case have been examined previously in Mellin space in \cite{Alday:2024srr}. However, for the giant graviton case the Mellin space formalism is not fully available due to a technical issue \cite{giantgravitonlongpaper}. Therefore, this is a situation where the complementary position space approach is necessary and can be more useful.

In the flat-space limit, we should reproduce the 1-to-1 form factor of gravitons scattering with an extended $q$ dimensional brane. The flat-space form factor can be computed from string theory and is known in the literature for general $q$ \cite{Klebanov:1995ni,Hashimoto:1996bf,Garousi:1996ad}. In this section, we will focus on the supergravity limit and the result can be obtained by further taking the limit of $\ell_s$ tending to zero. More precisely, the flat-space graviton form factor is given by (up to an overal numerical factor which we do not track)
\begin{equation}
    \mathcal{A}_{\rm flat}\propto \frac{1}{st}(sa_1-ta_2)\;,
\end{equation}
where $t=-S$, $s=-2Q$, and 
\begin{equation}
\begin{split}
a_1={}&{\rm Tr}(e_1\cdot D)p_1\cdot e_2\cdot p_1-p_1\cdot e_2\cdot D\cdot e_1\cdot p_2-2p_1\cdot e_2\cdot e_1 \cdot D \cdot p_1-p_1\cdot e_2 \cdot e_1 \cdot p_2\\
{}&+\frac{s}{2}{\rm Tr}(e_1 \cdot e_2)+(1\leftrightarrow 2)\;,\\
a_2={}&{\rm Tr}(e_1\cdot D)(p_1\cdot e_2\cdot D\cdot p_2+p_2\cdot D\cdot e_2\cdot p_1+p_2\cdot D\cdot e_2\cdot D\cdot p_2)\\
{}&+p_1\cdot D\cdot e_1\cdot D\cdot e_2\cdot D\cdot p_2-p_2\cdot D\cdot e_2\cdot e_1\cdot D\cdot p_1+\frac{s}{2}{\rm Tr}(e_1\cdot D\cdot e_2\cdot D)\\
{}&-\frac{s}{2}{\rm Tr}(e_1\cdot e_2)-\frac{s+t}{2}{\rm Tr}(e_1\cdot D){\rm Tr}(e_2\cdot D)+(1\leftrightarrow 2)\;.
\end{split}
\end{equation}
Here $e_{i,\mu\nu}$ are polarization tensors of the gravitons. The orientation of the brane in the total spacetime is characterized by a diagonal rank-2 tensor
\begin{equation}
    D^\mu{}_\nu={\rm diag}(1,\ldots,1,-1,\ldots,-1)\;,
\end{equation}
where the first $q$ elements are $+1$ and the remaining elements are $-1$. 

Before we start considering each specific case, let us make a few general comments. To make connections between flat-space amplitudes and holographic correlators, we need to identify the various quantities on the two sides. In flat space, the scattering states are spin-2 gravitons. However, in defect CFTs we study correlation functions of scalar operators dual to supergravitons. In all the cases which we will consider, these operators transform in the rank-$k$ symmetric-traceless representations of a certain R-symmetry group $SO(n)$. The R-symmetry information is captured by introducing $n$ dimensional null vectors 
\begin{equation}
    O_k(x,u)=O^{I_1\ldots I_k}_k(x)u^{I_1}\ldots u^{I_k}\;,
\end{equation}
where $u^I\cdot u^I=0$. In the flat-space limit, the R-symmetry polarization vector is related to the graviton polarization tensor via 
\begin{equation}
    e_{i,\mu\nu}=u_{i,\mu}u_{i,\nu}\;.
\end{equation}
Here it is understood that that the vectors $u$ are uplifted into the bigger 10d or 11d space by adding zeros. Moreover, because $u$ points in the internal dimensions orthogonal to the AdS factor, the R-symmetry polarization vectors are perpendicular to the flat-space particle momenta \cite{Chester:2018aca,Alday:2021odx}
\begin{equation}
    u_i\cdot p_j=0\;,
\end{equation}
for any pairs of $i$ and $j$. This implies
\begin{equation}
    p_i\cdot e_j=0\;,
\end{equation}
and simplifies $a_1$ and $a_2$ into
\begin{equation}
\begin{split}
a_1={}&
s(u_1\cdot u_2)^2\;,\\
a_2={}&(p_2\cdot D\cdot e_2\cdot D\cdot p_2){\rm Tr}(e_1\cdot D)+p_1\cdot D\cdot e_1\cdot D\cdot e_2\cdot D\cdot p_2-p_2\cdot D\cdot e_2\cdot e_1\cdot D\cdot p_1\\
{}&+\frac{s}{2}{\rm Tr}(e_1\cdot D\cdot e_2\cdot D)-\frac{s}{2}{\rm Tr}(e_1\cdot e_2)-\frac{s+t}{2}{\rm Tr}(e_1\cdot D){\rm Tr}(e_2\cdot D)+(1\leftrightarrow 2)\;.
\end{split}
\end{equation}
The flat-space tensor $D$ can also be related to the quantities in defect CFTs. But the precise relation depends on the specific setup and the details will be discussed case by case.

\subsection{Wilson loops in 4d $\mathcal{N}=4$ SYM}
Let us first consider $\frac{1}{2}$-BPS Wilson loops in 4d $\mathcal{N}=4$ SYM. Two-point functions of $\frac{1}{2}$-BPS operators inserted away from the Wilson loop have been computed in the supergravity limit at tree level in \cite{Barrat:2021yvp,Barrat:2022psm,Gimenez-Grau:2023fcy}. As mentioned, the flat-space limit has been analyzed in Mellin space in \cite{Alday:2024srr}. Here we will repeat the analysis of this limit in position space. 

For the purpose of demonstration, it is sufficient to focus on the simplest case where the operators have the lowest KK level $k_1=k_2=2$. In position space, the result can be written as a sum of Witten diagrams \cite{Gimenez-Grau:2023fcy}
\begin{equation}
\mathcal{F}_{\rm sugra}=\mathcal{N}_{\rm WL}\left(\sigma\left(1-\frac{\sigma}{6}\right)\mathcal{P}^{2,0}_{22}+\frac{\sigma^2}{360}\mathcal{P}^{4,2}_{22}+(\sigma-1)\widehat{\mathcal{W}}^{1,0}_{22}+\frac{1}{4}\widehat{\mathcal{W}}^{2,1}_{22}+\frac{1-2\sigma}{\pi}\mathcal{W}^{\rm contact}_{22}\right)\;,
\end{equation}
where $\mathcal{N}_{\rm WL}\sim \sqrt{\lambda}/N^2$ is an overall coefficient. In this case, the Wilson loop sits at a point in $S^5$ which can be characterized by a fixed six dimensional vector $\theta$ satisfying $\theta\cdot \theta=1$. This breaks the R-symmetry from $SO(6)$ to $SO(5)$ and allows us to define an R-symmetry cross ratio\footnote{Let us warn the reader about a slight abuse of notation in this section where we recycle the variables to denote similar quantities in different setups. However, this should not cause confusions and the reader should look for the definitions within each subsection.}
\begin{equation}\label{defsigma}
    \sigma=\frac{u_1\cdot u_2}{(u_1\cdot \theta)(u_2\cdot \theta)}\;.
\end{equation}
To more compactly write down the Witten diagrams, it is useful to use the $x$, $\bar{x}$ variables introduced in (\ref{defzzb}) and  define
\begin{equation}
    r=\sqrt{x\bar{x}}\;.
\end{equation}
Note that in the flat-space limit, we need to take $r\to-1$. The bulk channel exchange Witten diagrams $\mathcal{P}^{\Delta,\ell}_{\Delta_1,\Delta_2}$ are\footnote{More precisely, they are Polyakov-Regge blocks where the exchange Witten diagrams are combined with contact Witten diagrams to improve the Regge behevior \cite{Mazac:2018biw,Mazac:2019shk,Gimenez-Grau:2023fcy}.}
\begin{equation}
\begin{split}
\mathcal{P}^{2,0}_{22}={}&\frac{2 r^2 \log r}{(r-1) (r+1) \left(r^2-r \chi +1\right)}\;,\\
\mathcal{P}^{4,2}_{22}={}&\frac{60 r^2 \left(-3 r^4+2 \left(r^4+4 r^2+1\right) \log (r)+3\right)}{\left(r^2-1\right)^3 \left(r^2-r \chi +1\right)}\;,
\end{split}
\end{equation}
and the defect channel exchange Witten diagrams 
 $\widehat{\mathcal{W}}^{\widehat{\Delta},s}_{\Delta_1\Delta_2}$ read
 \begin{equation}
\begin{split}
{}&\widehat{\mathcal{W}}^{1,0}_{22}=\log (r+1)-\frac{r^2 \log r}{r^2-1}\;,\\
{}&\widehat{\mathcal{W}}^{2,1}_{22}=\chi  \left(\frac{-2 r^4+r^3+4 r^2+r-2}{2 \left(r^2-1\right)^2}-\frac{\left(r^4-2 r^2+3\right) r^3 \log r}{\left(r^2-1\right)^3}+\left(r+\frac{1}{r}\right) \log (r+1)\right)\;.
\end{split}
\end{equation}
The contact Witten diagram in these variables is
\begin{equation}
\mathcal{W}^{\rm contact}_{22}=\frac{\pi  r^2 \left(-r^2+\left(r^2+1\right) \log (r)+1\right)}{2 \left(r^2-1\right)^3}\;.
\end{equation}
It is easy to see that under analytic continuation $r$ starts from the positive real axis and traces out a trajectory in the upper half plane before intersecting the negative real axis and ends slightly below it. To obtain the flat-space limit, we expand the Witten diagrams around $r=-1$ and take only the leading singularity. We find 
\begin{equation}
        \mathcal{P}^{4,2}_{22}\approx -\frac{180\sqrt{2}i\pi R^3}{\ell_s^3\eta^3(2+\chi)}\;,\quad \widehat{\mathcal{W}}^{2,1}_{22}\approx\frac{i\pi R^3\chi}{\sqrt{2}\ell_s^3\eta^3}\;,\quad \mathcal{W}^{\rm contact}_{22}\approx -\frac{i\pi R^3\chi}{2\sqrt{2}\ell_s^3\eta^3}\;,
\end{equation}
where we have used $\eta\approx \sqrt{2}(\frac{R}{\ell_s})(1+r)$ in the limit $r\to -1$. All the other Witten diagrams have singularities weaker than $\eta^{-3}$ and are therefore subleading. Substituting this scaling limit into the flat-space limit formula (\ref{eq:flatformulaF}), we get up to an overall constant  
\begin{equation}
    \mathcal{F}_{\text{flat, WL}}\approx \frac{(S+Q\sigma)^2}{QS}\;.
\end{equation}

On the other hand, the flat-space amplitude in the supergravity limit is \cite{Alday:2024srr}
\begin{equation}
\mathcal{A}_{\rm WL}\propto \frac{(s(u_1\cdot u_2)+2t(u_1\cdot \theta)(u_2\cdot\theta))^2}{st}\;.
\end{equation}
Here we have used 
\begin{equation}\label{DWL}
D_{\mu\nu}=2(\theta_\mu\theta_\nu+\zeta_\mu\zeta_\nu)-\eta_{\mu\nu}\;,
\end{equation}
where $\zeta$ is another unit vector orthogonal to $\theta$ and satisfies $u_i\cdot \zeta=0$. We find that the flat-space amplitude is precisely reproduced by our formula.

\subsection{Surface defects in 6d $(2,0)$ theories}
Let us now consider 6d $(2,0)$ theory in the presence of an $\frac{1}{2}$-BPS surface defect. In the holographic limit, the surface defect is realized as an M2 brane occupying an AdS$_3$ subspace embedded in AdS$_7\times$S$^4$. The tree-level defect correlators of two $\frac{1}{2}$-BPS operators inserted away from the surface defect have been systematically computed in \cite{Chen:2023yvw}.\footnote{The simplest correlator with the lowest KK modes with $k_1=k_2=2$ was first obtained in \cite{Meneghelli:2022gps}.} Based on the tree-level results, the one-loop correlator was computed in \cite{Chen:2024orp} using unitarity methods and the flat-space limit of this one-loop result was examined in \cite{Alday:2024srr}. In this subsection, we consider the flat-space limit of the tree-level result in position space, focusing on the lowest KK modes.\footnote{It would be interesting to also examine the flat-space limit of the one-loop correlator using the position space formula. However, this is currently not possible because the one-loop correlator is not known in position space.} 

These lowest KK modes are operators with conformal dimension $4$ and transform in the rank-$2$ symmetric traceless representation of the $SO(5)$ R-symmetry. Thanks to superconformal symmetry, the defect two-point function can be written in the following form \cite{Meneghelli:2022gps}
\begin{equation}
    G=\frac{(u_1\cdot \theta)^2(u_2\cdot \theta)^2}{|x_1^i|^4|x_2^i|^4}(\mathcal{G}_{\rm prot}+R^{(2)}\, \mathcal{H})\;.
\end{equation}
Here $\mathcal{G}_{\rm prot}$ is the protected part of the correlator
\begin{equation}
    \mathcal{G}_{\rm prot}=\frac{r^2 \sigma }{\left(r^2-r \chi +1\right)^2}\;,
\end{equation}
and $\mathcal{H}$ is the reduced correlator which contains dynamical information
\begin{equation}
\begin{split}
    \mathcal{H}={}&\frac{r^4}{2 \left(r^2-1\right)^7 \left(r^2-r \chi +1\right)^2}\times\bigg(\left(r^2-1\right) \big(-2 r^8+r^7 \chi -38 r^6+29 r^5 \chi -40 r^4+29 r^3 \chi\\
    {}& -38 r^2+r \chi -2\big)+12 r^2 \left(3 r^6+7 r^4+7 r^2-2 \left(r^4+3 r^2+1\right) r \chi +3\right) \log (r)\bigg)\;.
\end{split}
\end{equation}
We have also omitted an overall coefficient of order $1/N^2$ which we do not keep track of in this section. Because the surface defect also sits at a point in the internal S$^4$, we can  represent it using a fixed five-dimensional vector $\theta$ and define the R-symmetry cross ratio $\sigma$ in the same way as in (\ref{defsigma}). The factor $R^{(2)}$ is determined by superconformal symmetry to be
\begin{equation}
    R^{(2)}=\frac{(x-\omega)(\bar{x}-\omega)(x-\omega^{-1})(\bar{x}-\omega^{-1})}{x\bar{x}}\;,
\end{equation}
where 
\begin{equation}
    \sigma=-\frac{(1-\omega)^2}{2\omega}\;.
\end{equation}
We now examine the behavior of the correlator at the scaling limit. It is straightforward to find
\begin{equation}\label{Hsurfacesing}
    \mathcal{H}\approx-\frac{15\pi i R^7}{2\sqrt{2}\ell_s^7\eta^7(2+\chi)}\;,
\end{equation}
while the protected part $\mathcal{G}_{\rm prot}$ is regular. Therefore, the flat-space amplitude only receives contribution from the reduced correlator. This is in a way similar to the phenomenon observed in \cite{Rastelli:2016nze,Rastelli:2017udc} that only the reduced correlator contributes to the Mellin amplitude. Substituting (\ref{Hsurfacesing}) into the flat-space formula, we find 
\begin{equation}
\mathcal{F}_{\text{flat, surface}}\propto \frac{(S+Q\sigma)^2}{QS}\;,    
\end{equation}
where the factor $R^{(2)}$ contributes a factor $4(S+Q\sigma)^2/Q^2$.

Let us now consider the form factor in flat space. The tensor $D_{\mu\nu}$ can be related to the CFT variables in a way similar to (\ref{DWL})
\begin{equation}
D_{\mu\nu}=2(\theta_\mu\theta_\nu+\zeta_{1,\mu}\zeta_{1,\nu}+\zeta_{2,\mu}\zeta_{2,\nu})-\eta_{\mu\nu}\;,
\end{equation}
where we need an extra $\zeta$ vector since the defect is three dimensional in flat space. These vectors satisfy the orthogonality condition
\begin{equation}
    \theta\cdot \zeta_a=0\;, \quad \zeta_1\cdot \zeta_2=0\;,\quad u_i\cdot \zeta_a=0\;.
\end{equation}
It is easy to see that the flat-space amplitude the same as in the Wilson loop case
\begin{equation}
\mathcal{A}_{\rm surface}\propto \frac{(s(u_1\cdot u_2)+2t(u_1\cdot \theta)(u_2\cdot\theta))^2}{st}\;,
\end{equation}
and agrees with what we get from the flat-space limit formula.

\subsection{Giant gravitons in 4d $\mathcal{N}=4$ SYM}
Finally, we consider four-point functions in 4d $\mathcal{N}=4$ SYM which have two maximal giant gravitons and two light supergravitons. As was shown in \cite{Chen:2025yxg}, these four-point functions are most naturally viewed as two-point functions of supergravitons in the presence of a defect created by the two heavy giant gravitons. The defect has a seemingly exotic dimension of zero but this fact is easy to understand from the holographic perspective. The giant graviton operators are dual to heavy states in AdS which trace out a one dimensional geodesic line between the two insertion points on the boundary. The defect in the boundary CFT has one dimension fewer and therefore is zero dimensional.

We will also focus on the simplest case where the supergravitons have the lowest KK weights. We insert the two supergravitons at $x_1$, $x_2$ and their R-symmetry polarization vectors are $u_1$, $u_2$. The giant gravitons are inserted at $x_3$, $x_4$ with R-symmetry polarizations $u_3$, $u_4$. The defect two-point function is formed by the dividing the four-point function by the giant graviton two-point function \cite{Chen:2025yxg}. Superconformal symmetry dictates it to take the form
\begin{equation}\label{Ggiantgravitonsplit}
    G=G_{\rm free}+\frac{(u_1\cdot u_2)^2x_{13}^4x_{24}^4}{x_{34}^4} R^{(4)}\, H\;.
\end{equation}
The factor $R^{(4)}$ is given by
\begin{equation}
R^{(4)}=(1-z\alpha)(1-z\bar{\alpha})(1-\bar{z}\alpha)(1-\bar{z}\bar{\alpha})\;,
\end{equation}
where we have used the four-point function cross ratios
\begin{equation}\label{crossratios4pt}
\begin{split}
    {}&U=\frac{x_{12}^2x_{34}^2}{x_{13}^2x_{24}^2}=z\bar{z}\;,\quad\quad\quad\quad\quad V=\frac{x_{14}^2x_{23}^2}{x_{13}^2x_{24}^2}=(1-z)(1-\bar{z})\;,\\ {}&\sigma=\frac{(u_1\cdot u_3)(u_2\cdot u_4)}{(u_1\cdot u_2)(u_3\cdot u_4)}=\alpha\bar{\alpha}\;,\quad\quad \tau=\frac{(u_1\cdot u_4)(u_2\cdot u_3)}{(u_1\cdot u_2)(u_3\cdot u_4)}=(1-\alpha)(1-\bar{\alpha})\;.
\end{split}
\end{equation}
Here let us remind the reader that the four-point function cross ratios $z$, $\bar{z}$ are related to the defect two-point function cross ratios $x$, $\bar{x}$ via (\ref{zzbandxxb}). As was shown in \cite{Chen:2025yxg}, the leading nontrivial large $N$ contribution to the two-pint function comes from tree-level supergravity Witten diagrams and the combination of these diagrams is completely fixed by superconformal symmetry. The result can then be written in the form of (\ref{Ggiantgravitonsplit}) where the free correlator is \cite{Chen:2025yxg}
\begin{equation}
    G_{\rm free}=\frac{2(u_1\cdot u_2)^2x_{34}^4(\tau-\sigma\tau U+\sigma V)}{x_{13}^2x_{14}^2x_{23}^2x_{24}^2U}\;,
\end{equation}
and the reduced correlator is
\begin{equation}
    H=\frac{2 x_{34}^8 V(V^2-2V\log V-1)}{ x_{13}^4x_{14}^4x_{23}^4x_{24}^4U(1-V)^3}\;.
\end{equation}
Compared to the previous two cases, a special feature of these giant graviton correlators is that the full correlator $G$ does not admit a defect Mellin representation. This is because the giant graviton correlators contain exchange Witten diagrams of vector fields in the bulk channel. Such diagrams do not have a representation in the Mellin formalism and are absent in the two other setups \cite{giantgravitonlongpaper}. Therefore, the giant graviton correlators are a place where our position space formula has an advantage over the Mellin space one. Taking the flat-space limit is similar to the previous case of surface defect in the 6d $(2,0)$ theory. Only the reduced correlator contributes to the scaling limit and we get
\begin{equation}
    H\approx\frac{\sqrt{2}\pi i x_{34}^8 R^3}{ x_{13}^4x_{14}^4x_{23}^4x_{24}^4\ell_s^3\eta^3(2+\chi)}\;.
\end{equation}
The flat-space limit formula then gives
\begin{equation}\label{Fflatgiant}
    \mathcal{F}_{\text{flat, giant graviton}}\approx \frac{(Q+2S(1-\alpha)\alpha)(Q+2S(1-\bar{\alpha})\bar{\alpha})}{QS}\;.
\end{equation}

To check our formula, we also need to express the 10d flat-space tensor $D$ in terms of the CFT variables. This tensor is given by the following combination
\begin{equation}\label{defDgiantgraviton}
    D_{\mu\nu}=2\mathbb{M}_{\mu\nu}-\eta_{\mu\nu}\;,
\end{equation}
where $\mathbb{M}$ is a projector introduced in \cite{Chen:2025yxg} to capture R-symmetry of the giant graviton defect
\begin{equation}
    \mathbb{M}_{\mu\nu}=\mathbb{I}_{\mu\nu}^{\rm R}-\frac{u_{3,\mu}u_{4,\nu}+u_{4,\mu}u_{3,\nu}}{u_3\cdot u_4}\;.
\end{equation}
Here $\mathbb{I}^{\rm R}$ is a $6\times 6$ identity matrix in the subspace in which the  R-symmetry polarization vectors live. But we need to embed $\mathbb{M}$ in the $10\times 10$ matrix by extending the index ranges and setting to zeros the new matrix elements.\footnote{Before taking the flat-space limit, $\mathbb{M}$ projects only to a three dimensional S$^3$ inside of S$^5$ because the six dimensional embedding vector $y$ is further constrained by the condition $y\cdot y=1$. Therefore, in \cite{Chen:2025yxg} it was combined with an extra projector in the orthogonal AdS factor. The other projector supplements another $\mathbb{R}$ to make up the four dimensional defect in AdS$_5\times$S$^5$ which is responsible for the emergence of the hidden conformal symmetry. The situation in flat space is a bit different. As we will see below, $\mathbb{M}$ alone already projects to the four dimensional subspace occupied by the flat-space defect.} R-symmetry allows us to freely choose $u_3$ and $u_4$. Let us pick the first six components to correspond to the original six dimensional R-symmetry vectors, a convenient gauge is
\begin{equation}
    \begin{split}
        u_3={}&(0,0,0,0,1,i,0,0,0,0)\;,\\
        u_4={}&(0,0,0,0,1,-i,0,0,0,0)\;.
    \end{split}
\end{equation}
This gives us the diagonal tensor
\begin{equation}
    D={\rm diag}(1,1,1,1,-1,-1,-1,-1,-1,-1)\;.
\end{equation}
Because the flat-space momenta are orthogonal to the R-symmetry polarization vectors, we have
\begin{equation}
    p_i\cdot D\cdot e_j=0\;.
\end{equation}
Further using the explicit form of the $D$ tensor (\ref{defDgiantgraviton}), we get
\begin{equation}
    a_2=u_{12}^2\left(s((2(\sigma+\tau)-1)^2-1)-16\sigma\tau(s+t)\right)\;.
\end{equation}
The flat-space form factor is then given by
\begin{equation}
    \mathcal{A}_{\text{giant graviton}}=\frac{(s+4(1-\alpha)\alpha t)(s+4(1-\bar{\alpha})\bar{\alpha} t)}{st}\;,
\end{equation}
which agrees precisely with the result (\ref{Fflatgiant}) from taking the flat-space limit.

\section{Discussions}\label{Sec:discussions}
In this paper, we derived a position space formula for taking the large AdS radius limit of holographic defect correlators. The formula involves a special scaling limit of defect two-point functions in the Lorentzian regime and gives the flat-space form factor by integrating the correlator against a Bessel function kernel. Our formula is valid for a wide range of defect dimensions $0\leq p\leq d-1$, including the case of $p=0$ where a Mellin space representation is not always available. When defect Mellin amplitudes can be defined, we also proved that our result is equivalent to the flat-space limit formula recently conjectured in Mellin space \cite{Alday:2024srr}. These results are similar to the ones in defect-free four-point functions \cite{Okuda:2010ym, Penedones:2010ue}, but extend them in interesting ways.

As nontrivial checks for our formula, we studied several known examples in the tree-level supergravity limit in holographic defect CFTs. However, it would be more interesting to use the flat-space limit formula as a tool to obtain new results.
An important application is to study the stringy and M-theory corrections to defect correlators beyond the supergravity approximation, as has been initiated in Mellin space in \cite{Alday:2024srr}. Another interesting direction to explore is to consider defect correlators at loop levels \cite{Chen:2024orp}. Having a complementary formula in position space may offer some technical advantage to these explorations. It should also be noted that in the latter case no position space method has so far been established. Our result will provide an important ingredient in the development of bootstrap techniques to compute loop-level defect holographic correlators in position space. It would also be very interesting to see if one can develop CFT methods to directly study the scaling limit of the  correlators relevant for the flat-space limit, without first computing the full defect two-point functions.

It should also be noted that we derived our result from a probe brane setup. It would be important to see if the formula can be applied more generally to ``non-perturbative'' examples where the defect back-reacts to the bulk geometry, and to understand the modifications needed if there are any. It would also be beneficial to consider in detail the special case of $p=-1$ dimensional defects which correspond to a point in AdS. The point-like nature of these AdS defects makes a separate analysis seem necessary and it is not immediately clear whether our formula already covers this case by just taking a special limit. Let us also point out that for such defects there is no known prescription for taking the flat-space limit even in Mellin space. In fact, the Mellin representation does not appear to be a convenient formalism, at least based on the explicit results in the supergravity limit \cite{Zhou:2024ekb}. However, the existence of a useful flat-space limit formula in position space seems reasonable to expect. Such a formula for $p=-1$ dimensional defects  would play a useful role in studying holographic CFTs on real projective space \cite{Verlinde:2015qfa,Nakayama:2015mva,Giombi:2020xah,Zhou:2024ekb}.

Finally, we can also try to extend the analysis to higher-point defect correlators which has not been explored very much \cite{Chen:2023oax}. In Mellin space, it was found that the prescription takes a universal form for all multiplicities \cite{Alday:2024srr}. It would be interesting to find its counterpart in position space. Let us also note that in this work we have  assumed that the particles are massless in the flat-space limit. It would be useful to extend the discussion to the most general case where some operators have dimensions $\Delta\sim R/\ell_s$ and become massive in flat space.

\acknowledgments
We thank Joao Penedones for discussions. Y.T. is supported by the National Key R\&D Program of China (NO. 2020YFA0713000 and NO. 124B2094). The work of X.Z. is supported by the NSFC Grant No. 12275273, funds from Chinese Academy of Sciences, University of Chinese Academy of Sciences, and the Kavli Institute for Theoretical Sciences. The work of X.Z. is also supported by the Xiaomi Foundation.

\bibliography{refs} 

\providecommand{\href}[2]{#2}\begingroup\raggedright\begin{thebibliography}{10}

\bibitem{Rastelli:2016nze}
L.~Rastelli and X.~Zhou, ``{Mellin amplitudes for $AdS_5\times S^5$},''
  \href{http://dx.doi.org/10.1103/PhysRevLett.118.091602}{{\em Phys. Rev.
  Lett.} {\bfseries 118} no.~9, (2017) 091602},
\href{http://arxiv.org/abs/1608.06624}{{\ttfamily arXiv:1608.06624 [hep-th]}}.
%%CITATION = ARXIV:1608.06624;%%.

\bibitem{Rastelli:2017udc}
L.~Rastelli and X.~Zhou, ``{How to Succeed at Holographic Correlators Without
  Really Trying},'' \href{http://dx.doi.org/10.1007/JHEP04(2018)014}{{\em JHEP}
  {\bfseries 04} (2018) 014},
\href{http://arxiv.org/abs/1710.05923}{{\ttfamily arXiv:1710.05923 [hep-th]}}.
%%CITATION = ARXIV:1710.05923;%%.

\bibitem{Gimenez-Grau:2023fcy}
A.~Gimenez-Grau, ``{The Witten Diagram Bootstrap for Holographic Defects},''
  \href{http://arxiv.org/abs/2306.11896}{{\ttfamily arXiv:2306.11896
  [hep-th]}}.

\bibitem{Chen:2023yvw}
J.~Chen, A.~Gimenez-Grau, and X.~Zhou, ``{Defect two-point functions in 6D
  (2,0) theories},'' \href{http://dx.doi.org/10.1103/PhysRevD.109.L061903}{{\em
  Phys. Rev. D} {\bfseries 109} no.~6, (2024) L061903},
  \href{http://arxiv.org/abs/2310.19230}{{\ttfamily arXiv:2310.19230
  [hep-th]}}.

\bibitem{Chen:2024orp}
J.~Chen, A.~Gimenez-Grau, H.~Paul, and X.~Zhou, ``{Unitarity Method for
  Holographic Defects},'' \href{http://arxiv.org/abs/2406.13287}{{\ttfamily
  arXiv:2406.13287 [hep-th]}}.

\bibitem{Zhou:2024ekb}
X.~Zhou, ``{Correlators of N=4 Supersymmetric Yang-Mills Theory on Real
  Projective Space at Strong Coupling},''
  \href{http://dx.doi.org/10.1103/PhysRevLett.133.201602}{{\em Phys. Rev.
  Lett.} {\bfseries 133} no.~20, (2024) 201602},
  \href{http://arxiv.org/abs/2408.04926}{{\ttfamily arXiv:2408.04926
  [hep-th]}}.

\bibitem{Chen:2025yxg}
J.~Chen, Y.~Jiang, and X.~Zhou, ``{Giant Graviton Correlators as Defect
  Systems},'' \href{http://arxiv.org/abs/2503.22987}{{\ttfamily
  arXiv:2503.22987 [hep-th]}}.

\bibitem{Bissi:2022mrs}
A.~Bissi, A.~Sinha, and X.~Zhou, ``{Selected topics in analytic conformal
  bootstrap: A guided journey},''
  \href{http://dx.doi.org/10.1016/j.physrep.2022.09.004}{{\em Phys. Rept.}
  {\bfseries 991} (2022) 1--89},
  \href{http://arxiv.org/abs/2202.08475}{{\ttfamily arXiv:2202.08475
  [hep-th]}}.

\bibitem{Caron-Huot:2018kta}
S.~Caron-Huot and A.-K. Trinh, ``{All Tree-Level Correlators in
  AdS${}_5\times$S${}_5$ Supergravity: Hidden Ten-Dimensional Conformal
  Symmetry},'' \href{http://dx.doi.org/10.1007/JHEP01(2019)196}{{\em JHEP}
  {\bfseries 01} (2019) 196},
\href{http://arxiv.org/abs/1809.09173}{{\ttfamily arXiv:1809.09173 [hep-th]}}.
%%CITATION = ARXIV:1809.09173;%%.

\bibitem{Rastelli:2019gtj}
L.~Rastelli, K.~Roumpedakis, and X.~Zhou, ``{$\mathbf{AdS_3\times S^3}$
  Tree-Level Correlators: Hidden Six-Dimensional Conformal Symmetry},''
  \href{http://dx.doi.org/10.1007/JHEP10(2019)140}{{\em JHEP} {\bfseries 10}
  (2019) 140},
\href{http://arxiv.org/abs/1905.11983}{{\ttfamily arXiv:1905.11983 [hep-th]}}.
%%CITATION = ARXIV:1905.11983;%%.

\bibitem{Giusto:2020neo}
S.~Giusto, R.~Russo, A.~Tyukov, and C.~Wen, ``{The CFT$_6$ origin of all
  tree-level 4-point correlators in AdS$_3 \times S^3$},''
  \href{http://dx.doi.org/10.1140/epjc/s10052-020-8300-4}{{\em Eur. Phys. J. C}
  {\bfseries 80} no.~8, (2020) 736},
  \href{http://arxiv.org/abs/2005.08560}{{\ttfamily arXiv:2005.08560
  [hep-th]}}.

\bibitem{Behan:2021pzk}
C.~Behan, P.~Ferrero, and X.~Zhou, ``{More on holographic correlators: Twisted
  and dimensionally reduced structures},''
  \href{http://dx.doi.org/10.1007/JHEP04(2021)008}{{\em JHEP} {\bfseries 04}
  (2021) 008}, \href{http://arxiv.org/abs/2101.04114}{{\ttfamily
  arXiv:2101.04114 [hep-th]}}.

\bibitem{Alday:2021odx}
L.~F. Alday, C.~Behan, P.~Ferrero, and X.~Zhou, ``{Gluon Scattering in AdS from
  CFT},'' \href{http://dx.doi.org/10.1007/JHEP06(2021)020}{{\em JHEP}
  {\bfseries 06} (2021) 020}, \href{http://arxiv.org/abs/2103.15830}{{\ttfamily
  arXiv:2103.15830 [hep-th]}}.

\bibitem{Zhou:2021gnu}
X.~Zhou, ``{Double Copy Relation in AdS Space},''
  \href{http://dx.doi.org/10.1103/PhysRevLett.127.141601}{{\em Phys. Rev.
  Lett.} {\bfseries 127} no.~14, (2021) 141601},
  \href{http://arxiv.org/abs/2106.07651}{{\ttfamily arXiv:2106.07651
  [hep-th]}}.

\bibitem{Alday:2022lkk}
L.~F. Alday, V.~Gon\c{c}alves, and X.~Zhou, ``{Supersymmetric Five-Point Gluon
  Amplitudes in AdS Space},''
  \href{http://dx.doi.org/10.1103/PhysRevLett.128.161601}{{\em Phys. Rev.
  Lett.} {\bfseries 128} no.~16, (2022) 161601},
  \href{http://arxiv.org/abs/2201.04422}{{\ttfamily arXiv:2201.04422
  [hep-th]}}.

\bibitem{Aprile:2025kfk}
F.~Aprile, C.~Behan, R.~S. Pitombo, and M.~Santagata, ``{Simplicity of Mellin
  amplitudes for AdS$_3 \times$S$^3$},''
  \href{http://arxiv.org/abs/2504.19770}{{\ttfamily arXiv:2504.19770
  [hep-th]}}.

\bibitem{Pufu:2023vwo}
S.~S. Pufu, V.~A. Rodriguez, and Y.~Wang, ``{Scattering From $(p,q)$-Strings in
  $\text{AdS}_5 \times \text{S}^5$},''
  \href{http://arxiv.org/abs/2305.08297}{{\ttfamily arXiv:2305.08297
  [hep-th]}}.

\bibitem{Billo:2023ncz}
M.~Billo', F.~Galvagno, M.~Frau, and A.~Lerda, ``{Integrated correlators with a
  Wilson line in $ \mathcal{N} $ = 4 SYM},''
  \href{http://dx.doi.org/10.1007/JHEP12(2023)047}{{\em JHEP} {\bfseries 12}
  (2023) 047}, \href{http://arxiv.org/abs/2308.16575}{{\ttfamily
  arXiv:2308.16575 [hep-th]}}.

\bibitem{Dempsey:2024vkf}
R.~Dempsey, B.~Offertaler, S.~S. Pufu, and Y.~Wang, ``{Global Symmetry and
  Integral Constraint on Superconformal Lines in Four Dimensions},''
  \href{http://arxiv.org/abs/2405.10914}{{\ttfamily arXiv:2405.10914
  [hep-th]}}.

\bibitem{Billo:2024kri}
M.~Bill\`o, M.~Frau, F.~Galvagno, and A.~Lerda, ``{A note on integrated
  correlators with a Wilson line in $\mathcal{N}=4$ SYM},''
  \href{http://arxiv.org/abs/2405.10862}{{\ttfamily arXiv:2405.10862
  [hep-th]}}.

\bibitem{Brown:2024tru}
A.~Brown, F.~Galvagno, and C.~Wen, ``{Exact results for giant graviton
  four-point correlators},''
  \href{http://dx.doi.org/10.1007/JHEP07(2024)049}{{\em JHEP} {\bfseries 07}
  (2024) 049}, \href{http://arxiv.org/abs/2403.17263}{{\ttfamily
  arXiv:2403.17263 [hep-th]}}.

\bibitem{Alday:2023jdk}
L.~F. Alday, T.~Hansen, and J.~A. Silva, ``{Emergent Worldsheet for the AdS
  Virasoro-Shapiro Amplitude},''
  \href{http://dx.doi.org/10.1103/PhysRevLett.131.161603}{{\em Phys. Rev.
  Lett.} {\bfseries 131} no.~16, (2023) 161603},
  \href{http://arxiv.org/abs/2305.03593}{{\ttfamily arXiv:2305.03593
  [hep-th]}}.

\bibitem{Alday:2023mvu}
L.~F. Alday and T.~Hansen, ``{The AdS Virasoro-Shapiro amplitude},''
  \href{http://dx.doi.org/10.1007/JHEP10(2023)023}{{\em JHEP} {\bfseries 10}
  (2023) 023}, \href{http://arxiv.org/abs/2306.12786}{{\ttfamily
  arXiv:2306.12786 [hep-th]}}.

\bibitem{Alday:2023pzu}
L.~F. Alday, T.~Hansen, and M.~Nocchi, ``{High Energy String Scattering in
  AdS},'' \href{http://dx.doi.org/10.1007/JHEP02(2024)089}{{\em JHEP}
  {\bfseries 02} (2024) 089}, \href{http://arxiv.org/abs/2312.02261}{{\ttfamily
  arXiv:2312.02261 [hep-th]}}.

\bibitem{Alday:2024xpq}
L.~F. Alday, M.~Nocchi, C.~Virally, and X.~Zhou, ``{On the Regge behaviour of
  the AdS Virasoro-Shapiro amplitude},''
  \href{http://dx.doi.org/10.1007/JHEP04(2025)064}{{\em JHEP} {\bfseries 04}
  (2025) 064}, \href{http://arxiv.org/abs/2409.03695}{{\ttfamily
  arXiv:2409.03695 [hep-th]}}.

\bibitem{Alday:2024rjs}
L.~F. Alday, G.~Giribet, and T.~Hansen, ``{On the AdS$_{3}$ Virasoro-Shapiro
  amplitude},'' \href{http://dx.doi.org/10.1007/JHEP03(2025)002}{{\em JHEP}
  {\bfseries 03} (2025) 002}, \href{http://arxiv.org/abs/2412.05246}{{\ttfamily
  arXiv:2412.05246 [hep-th]}}.

\bibitem{Wang:2025pjo}
B.~Wang, D.~Wu, and E.~Y. Yuan, ``{The Kaluza-Klein AdS Virasoro-Shapiro
  Amplitude near Flat Space},''
  \href{http://arxiv.org/abs/2503.01964}{{\ttfamily arXiv:2503.01964
  [hep-th]}}.

\bibitem{Alday:2025bjp}
L.~F. Alday, M.~Nocchi, and A.~S. Sangar{\'e}, ``{Stringy KLT Relations on
  $AdS$},'' \href{http://arxiv.org/abs/2504.19973}{{\ttfamily arXiv:2504.19973
  [hep-th]}}.

\bibitem{Okuda:2010ym}
T.~Okuda and J.~Penedones, ``{String scattering in flat space and a scaling
  limit of Yang-Mills correlators},''
  \href{http://dx.doi.org/10.1103/PhysRevD.83.086001}{{\em Phys. Rev. D}
  {\bfseries 83} (2011) 086001},
  \href{http://arxiv.org/abs/1002.2641}{{\ttfamily arXiv:1002.2641 [hep-th]}}.

\bibitem{Penedones:2010ue}
J.~Penedones, ``{Writing CFT correlation functions as AdS scattering
  amplitudes},'' \href{http://dx.doi.org/10.1007/JHEP03(2011)025}{{\em JHEP}
  {\bfseries 03} (2011) 025},
\href{http://arxiv.org/abs/1011.1485}{{\ttfamily arXiv:1011.1485 [hep-th]}}.
%%CITATION = ARXIV:1011.1485;%%.

\bibitem{Alday:2024srr}
L.~F. Alday and X.~Zhou, ``{Flat-space limit of defect correlators and stringy
  AdS form factors},'' \href{http://dx.doi.org/10.1007/JHEP03(2025)182}{{\em
  JHEP} {\bfseries 03} (2025) 182},
  \href{http://arxiv.org/abs/2411.04378}{{\ttfamily arXiv:2411.04378
  [hep-th]}}.

\bibitem{Billo:2016cpy}
M.~Bill\`o, V.~Gon\c{c}alves, E.~Lauria, and M.~Meineri, ``{Defects in
  conformal field theory},''
  \href{http://dx.doi.org/10.1007/JHEP04(2016)091}{{\em JHEP} {\bfseries 04}
  (2016) 091}, \href{http://arxiv.org/abs/1601.02883}{{\ttfamily
  arXiv:1601.02883 [hep-th]}}.

\bibitem{Mazac:2018biw}
D.~Mazac, L.~Rastelli, and X.~Zhou, ``{An analytic approach to BCFT$_{d}$},''
  \href{http://dx.doi.org/10.1007/JHEP12(2019)004}{{\em JHEP} {\bfseries 12}
  (2019) 004}, \href{http://arxiv.org/abs/1812.09314}{{\ttfamily
  arXiv:1812.09314 [hep-th]}}.

\bibitem{Gary:2009ae}
M.~Gary, S.~B. Giddings, and J.~Penedones, ``{Local bulk S-matrix elements and
  CFT singularities},''
  \href{http://dx.doi.org/10.1103/PhysRevD.80.085005}{{\em Phys. Rev. D}
  {\bfseries 80} (2009) 085005},
  \href{http://arxiv.org/abs/0903.4437}{{\ttfamily arXiv:0903.4437 [hep-th]}}.

\bibitem{Rastelli:2017ecj}
L.~Rastelli and X.~Zhou, ``{The Mellin Formalism for Boundary CFT$_d$},''
  \href{http://dx.doi.org/10.1007/JHEP10(2017)146}{{\em JHEP} {\bfseries 10}
  (2017) 146},
\href{http://arxiv.org/abs/1705.05362}{{\ttfamily arXiv:1705.05362 [hep-th]}}.
%%CITATION = ARXIV:1705.05362;%%.

\bibitem{Mack:2009mi}
G.~Mack, ``{D-independent representation of Conformal Field Theories in D
  dimensions via transformation to auxiliary Dual Resonance Models. Scalar
  amplitudes},''
\href{http://arxiv.org/abs/0907.2407}{{\ttfamily arXiv:0907.2407 [hep-th]}}.
%%CITATION = ARXIV:0907.2407;%%.

\bibitem{Goncalves:2018fwx}
V.~Goncalves and G.~Itsios, ``{A note on defect Mellin amplitudes},''
  \href{http://arxiv.org/abs/1803.06721}{{\ttfamily arXiv:1803.06721
  [hep-th]}}.

\bibitem{giantgravitonlongpaper}
J.~Chen, Y.~Jiang, and X.~Zhou {\em In preparation.} .

\bibitem{Klebanov:1995ni}
I.~R. Klebanov and L.~Thorlacius, ``{The Size of p-branes},''
  \href{http://dx.doi.org/10.1016/0370-2693(95)01576-0}{{\em Phys. Lett. B}
  {\bfseries 371} (1996) 51--56},
  \href{http://arxiv.org/abs/hep-th/9510200}{{\ttfamily arXiv:hep-th/9510200}}.

\bibitem{Hashimoto:1996bf}
A.~Hashimoto and I.~R. Klebanov, ``{Scattering of strings from D-branes},''
  \href{http://dx.doi.org/10.1016/S0920-5632(97)00074-1}{{\em Nucl. Phys. B
  Proc. Suppl.} {\bfseries 55} (1997) 118--133},
  \href{http://arxiv.org/abs/hep-th/9611214}{{\ttfamily arXiv:hep-th/9611214}}.

\bibitem{Garousi:1996ad}
M.~R. Garousi and R.~C. Myers, ``{Superstring scattering from D-branes},''
  \href{http://dx.doi.org/10.1016/0550-3213(96)00316-1}{{\em Nucl. Phys. B}
  {\bfseries 475} (1996) 193--224},
  \href{http://arxiv.org/abs/hep-th/9603194}{{\ttfamily arXiv:hep-th/9603194}}.

\bibitem{Chester:2018aca}
S.~M. Chester, S.~S. Pufu, and X.~Yin, ``{The M-Theory S-Matrix From ABJM:
  Beyond 11D Supergravity},''
  \href{http://dx.doi.org/10.1007/JHEP08(2018)115}{{\em JHEP} {\bfseries 08}
  (2018) 115},
\href{http://arxiv.org/abs/1804.00949}{{\ttfamily arXiv:1804.00949 [hep-th]}}.
%%CITATION = ARXIV:1804.00949;%%.

\bibitem{Barrat:2021yvp}
J.~Barrat, A.~Gimenez-Grau, and P.~Liendo, ``{Bootstrapping holographic defect
  correlators in $ \mathcal{N} $ = 4 super Yang-Mills},''
  \href{http://dx.doi.org/10.1007/JHEP04(2022)093}{{\em JHEP} {\bfseries 04}
  (2022) 093}, \href{http://arxiv.org/abs/2108.13432}{{\ttfamily
  arXiv:2108.13432 [hep-th]}}.

\bibitem{Barrat:2022psm}
J.~Barrat, A.~Gimenez-Grau, and P.~Liendo, ``{A dispersion relation for defect
  CFT},'' \href{http://dx.doi.org/10.1007/JHEP02(2023)255}{{\em JHEP}
  {\bfseries 02} (2023) 255}, \href{http://arxiv.org/abs/2205.09765}{{\ttfamily
  arXiv:2205.09765 [hep-th]}}.

\bibitem{Mazac:2019shk}
D.~Maz{\'a}{\v c}, L.~Rastelli, and X.~Zhou, ``{A Basis of Analytic Functionals
  for CFTs in General Dimension},''
  \href{http://arxiv.org/abs/1910.12855}{{\ttfamily arXiv:1910.12855
  [hep-th]}}.

\bibitem{Meneghelli:2022gps}
C.~Meneghelli and M.~Tr\'epanier, ``{Bootstrapping string dynamics in the 6d
  $\mathcal{N}$ = (2, 0) theories},''
  \href{http://dx.doi.org/10.1007/JHEP07(2023)165}{{\em JHEP} {\bfseries 07}
  (2023) 165}, \href{http://arxiv.org/abs/2212.05020}{{\ttfamily
  arXiv:2212.05020 [hep-th]}}.

\bibitem{Verlinde:2015qfa}
H.~Verlinde, ``{Poking Holes in AdS/CFT: Bulk Fields from Boundary States},''
  \href{http://arxiv.org/abs/1505.05069}{{\ttfamily arXiv:1505.05069
  [hep-th]}}.

\bibitem{Nakayama:2015mva}
Y.~Nakayama and H.~Ooguri, ``{Bulk Locality and Boundary Creating Operators},''
  \href{http://dx.doi.org/10.1007/JHEP10(2015)114}{{\em JHEP} {\bfseries 10}
  (2015) 114}, \href{http://arxiv.org/abs/1507.04130}{{\ttfamily
  arXiv:1507.04130 [hep-th]}}.

\bibitem{Giombi:2020xah}
S.~Giombi, H.~Khanchandani, and X.~Zhou, ``{Aspects of CFTs on Real Projective
  Space},'' \href{http://dx.doi.org/10.1088/1751-8121/abcf59}{{\em J. Phys. A}
  {\bfseries 54} no.~2, (2021) 024003},
  \href{http://arxiv.org/abs/2009.03290}{{\ttfamily arXiv:2009.03290
  [hep-th]}}.

\bibitem{Chen:2023oax}
J.~Chen and X.~Zhou, ``{Aspects of higher-point functions in BCFT$_{d}$},''
  \href{http://dx.doi.org/10.1007/JHEP09(2023)204}{{\em JHEP} {\bfseries 09}
  (2023) 204}, \href{http://arxiv.org/abs/2304.11799}{{\ttfamily
  arXiv:2304.11799 [hep-th]}}.

\end{thebibliography}\endgroup
\bibliographystyle{utphys}
\end{document}